\documentclass[12pt]{article}
\pdfoutput=1
\usepackage{jheppub}
\usepackage{amsmath} 
\usepackage{amssymb,amsfonts}
\usepackage[dvipsnames]{xcolor}
\usepackage{graphicx}
\usepackage[utf8]{inputenc}
\numberwithin{equation}{section}

\newcommand{\pa}{\partial}

\newcommand{\nn}{\nonumber}

\newcommand{\be} {\begin{equation}}
\newcommand{\ee} {\end{equation}}
\renewcommand{\l}{{\lambda}}
\title{Modulated instabilities and the AdS$_2$ point in dense holographic matter}
\preprint{APCTP Pre2024 - 009}
\author[a,b]{Jes\'us Cruz Rojas}
\affiliation[a]{Asia Pacific Center for Theoretical Physics, Pohang, 37673, Korea}
\affiliation[b]{Departamento de F\'isica de Altas Energ\'ias, Instituto de Ciencias Nucleares, Universidad Nacional Aut\'onoma de M\'exico, Apartado Postal 70-543, CDMX 04510, M\'exico}
\author[c]{Tuna Demircik }
\affiliation[c]{Institute for Theoretical Physics, Wroclaw University of Science and Technology, 50-370 Wroclaw, Poland}
\author[a,d]{Matti J\"arvinen}
\affiliation[d]{Department of Physics, Pohang University of Science and Technology, Pohang, 37673, Korea}

\emailAdd{jesus.cruz@correo.nucleares.unam.mx}
\emailAdd{tuna.demircik@pwr.edu.pl}
\emailAdd{matti.jarvinen@apctp.org}

\abstract{
We investigate fluctuations of hot and dense QCD plasma by using the gauge/gravity duality. To this end, we
carry out a comprehensive classification and analysis of quasinormal modes of charged black holes in the holographic V-QCD model. It turns out that the Chern-Simons term determined by the flavor anomalies of QCD is strong enough to drive a modulated instability. While such an instability is expected at high densities, we find that the unstable phase extends to surprisingly low densities and high temperatures, close to the region where data from lattice simulations is available. 
We also analyze the limit of small temperature which is 
controlled by a quantum critical AdS$_2$ point. We study in detail the signatures of the critical point in the quasinormal mode spectrum, focusing on the interplay between the hydrodynamic modes and other modes.}

\begin{document}

\maketitle

\section{Introduction and summary}

Recent observations of binary neutron star mergers by LIGO and Virgo interferometers have provided new constraints for the behavior of QCD matter at high densities. 
These results are complemented by ongoing  heavy ion collisions experiments at RHIC, i.e., the beam energy scan program, as well as future heavy-ion experiments at facilities such as FAIR, NICA and J-PARC, which will push the range of data towards higher densities. 
Thanks to this incoming new data, there has also been enhanced interest in theoretical studies of dense QCD recently. Obtaining sound theoretical predictions for dense matter is however known to be particularly challenging. This is because none of the usual first-principles methods (perturbation theory, lattice simulations, or effective theory) work reliably at the relevant densities, i.e. the highest densities reached in heavy-ion collisions or the densities in the cores of massive neutron stars~\cite{Brambilla:2014jmp}.

Consequently, the precise structure of the phase diagram of QCD remains largely unknown at high densities. It is expected to contain various (usually deconfined) ``exotic'' phases in this region.  Typical examples of such phases are different color superconducting phases~\cite{Bailin:1983bm,Alford:1997zt,Alford:2007xm} (including the color-flavor locked phase~\cite{Alford:1998mk}), quarkyonic phases~\cite{McLerran:2007qj}, and spatially modulated phases.  
As an example of the latter, in \cite{Alford:2000ze}, a  crystalline color superconductor phase of quark matter was suggested. As a different example, it was shown that at low temperature and large $N_c$, the ground state of QCD  is a chiral density wave \cite{Deryagin:1992rw, Bringoltz:2009ym}. Quarkyonic chiral density waves have also been suggested~\cite{Kojo:2009ha}. Moreover, chiral magnetic wave \cite{Kharzeev:2010gd} 
and chiral magnetic spiral \cite{Basar:2010zd} as well as (confined) chiral soliton lattice~\cite{Son:2007ny,Brauner:2016pko,Evans:2022hwr} phases can be induced by a magnetic field.

In the absence of precise first-principles methods, modeling and new techniques can be useful in order to improve the theoretical understanding of the phases of dense QCD. A possibility is to use the gauge-gravity duality, which maps strong coupling physics in field theory to problems in classical gravity in higher dimensions. 

The approaches for holographic QCD can be roughly divided into top-down and bottom-up
models. In the first case the models are based on concrete (typically ten dimensional) setups in string theory for which precise control of the gauge theory is possible.  The bottom-up models are, in contrast, constructed ``by hand'' with
some inspiration from the top-down models. 
While some guidance is provided by the global symmetries of QCD which should be respected by the duality, a precise control of the correspondence is lost since the gravity models are not dual to any explicitly known field theory, but instead involve parameters, which should be adjusted to obtain agreement with QCD physics. However this  also means that there is a great amount of freedom to do modifications, which may be necessary to model QCD efficiently, but are difficult to realize in the top-down framework. This is the approach we will follow in this article.

Within the gauge-gravity duality, several examples of condensed superconducting phases (see, e.g.,~\cite{Gubser:2008px,Gubser:2008zu,Gubser:2008wv,Hartnoll:2008vx,Horowitz:2008bn}) and modulated phases~\cite{Domokos:2007kt,Nakamura:2009tf,Ooguri:2010xs} are known.
In particular, spatial modulation can arise due to a Chern-Simons (CS) terms on the higher dimensional gravity side. In the context of QCD, such a CS term reflects the flavor anomalies of the field theory~\cite{Wess:1971yu,Witten:1983tw,Witten:1998qj}.
The presence of spatially modulated ground state typically implies an instability of the homogeneous state, which can be analyzed by studying the quasinormal modes (QNMs) of the dual geometry.
QNMs are linearized fluctuation modes of a black hole which assume the plane wave form $\sim \exp\left(-i\omega t +i\mathbf q \cdot \mathbf{x}\right)$, with a discrete spectrum of  complex valued frequencies. In the gauge theory side, QNMs correspond to the poles of the retarded Green's function of the dual operator  \cite{Son:2002sd,Nunez:2003eq,Horowitz:1999jd,Kovtun:2005ev}. For a more comprehensive information of QNMs, interested readers are referred to detailed reviews \cite{Berti:2009kk,Konoplya:2011qq,Kovtun:2005ev}. 
The presence of a mode with positive imaginary part of the frequency is an instability. If the instability only appears at nonzero spatial momentum $q=|\mathbf{q}|$, it is modulated.

In \cite{Nakamura:2009tf,Ooguri:2010kt,Ooguri:2010xs} it was shown that the Reissner-Nordstrom black hole in the five dimensional anti–de Sitter space coupled to the Maxwell theory and with a CS term is indeed unstable when the CS coupling is sufficiently large. The first instance of this type of instability was discovered in a chiral symmetry broken confined phase of a specific bottom-up phenomenological model \cite{Domokos:2007kt}. In this model, the CS term mixes vector and axial non-Abelian fields, and the resulting instability leads to the condensation of corresponding light mesons. This condensation breaks both translational and rotational invariance of the ground state at critical densities remarkably close to the nuclear saturation density $n_s\approx 0.16 \,\mathrm{fm}^{-3}$. The spatially modulated instability has also been observed in the top-down Witten-Sakai-Sugimoto (WSS) model, as discussed in \cite{Ooguri:2010xs, Chuang:2010ku}. However, this work explores the scenario in the chiral symmetric deconfined phase, i.e. the quark-gluon phase of the WSS model. Through their analysis, the authors propose a spatially modulated phase transition driven by the CS coupling of Abelian gauge fields. This phase transition occurs above a critical baryon density of $2.4$ fm$^{-3}$ for three flavors (equivalent to about $15n_s$) at a temperature of 150~MeV. 
The associated inhomogeneous field theory ground state is likely to accommodates modulated chiral  
currents.  
Subsequently, a similar instability in the confined phase of the WSS model at low temperature and finite axial chemical potential was identified in \cite{Bayona:2011ab}.

Spatially modulated instabilities similar to the ones discussed above are prevalent in dual descriptions of condensed matter systems, such as Fermi liquids \cite{Bergman:2011rf,Jokela:2012se,Iizuka:2013ag} and superconductors \cite{Donos:2011ff,Donos:2012gg,Donos:2011qt,Donos:2011pn,Donos:2012yu} (the construction of new striped ground states for superconductors can be found in \cite{Donos:2012gg,Donos:2012wi,Rozali:2013ama,Withers:2013kva,Donos:2013wia,Iizuka:2012pn,Rozali:2012es}). Certain factors can contribute to the stabilization of these ground states in the presence of a CS coupling. Examples include $R^2$ corrections \cite{Takeuchi:2011uk} and the inclusion of a magnetic field \cite{Ballon-Bayona:2012qnu,Jokela:2012vn}. Efforts have also been done to study the Hubbard model, which is used to describe the transition between conducting and insulating systems: In \cite{Fujita:2014mqa} a holographic construction of a large-$N$ Bose-Hubbard model is presented. 

It is also well known that in gravitational solutions dual to $n+1$ dimensional field theories at finite charge density, an AdS$_2 \times \mathbb{R}^n$ geometry appears in the extremal limit of black holes. The AdS$_2$ point is related to the emergence of a quantum critical region in the dual field theory~\cite{Liu:2009dm,Faulkner:2009wj,Iqbal:2011in}, with scaling symmetry in time, and controlled by a $(0+1)$-dimensional CFT at the IR.  The IR region is known to be sensitive to instabilities, such as the superconducting or modulated instabilities discussed above, due to a low Breitenlohner-Freedman (BF) bound of the AdS$_2$ space~\cite{Hartnoll:2008vx,Amado:2009ts}. 
This kind of quantum critical region also arises in holographic models for quark matter~\cite{Alho:2012mh,Alho:2013hsa}.

To describe the long wavelength dynamics of interacting quantum fields one can usually rely on hydrodynamics. 
At non-zero temperature quantum critical phases are found near quantum phase transitions, and in such regions one expects that transport properties are universal and are governed by scaling properties of the critical point.
The physics of hydrodynamics and near criticality can be concretely characterized in terms of QNMs. That is,  the QNMs with frequencies vanishing as momentum goes to zero are the hydrodynamic modes and these modes contain information about all the hydrodynamic transport coefficients. 
On the other hand, nonhydrodynamic modes appear as additional poles of the retarded Green’s function, and
provide information beyond the hydrodynamic description. They play a role in determining the applicability regime of hydrodynamics \cite{Heller:2013fn} and provide a description of the gauge theory plasma at late times \cite{Heller:2013oxa,Heller:2012km}. 

Hydrodynamics breaks down at sufficiently short scales set by the local equilibration time and length, since additional degrees of freedom start to play a more significant role.
In \cite{Arean:2020eus} the authors study the breakdown of hydrodynamics in certain low temperature states dual to black holes with nearly-extremal AdS$_2\times \mathbb{R}^2$ near-horizon metrics. The authors found that the breakdown of hydrodynamics is caused by modes associated to the AdS$_2$ region of the geometry, and as a consequence the critical frequency is set by data coming from the IR. 
The breakdown is characterized by a collision, parametrically close to the imaginary $\omega$ axis, between the lowest nonhydrodynamic mode and the hydrodynamic mode. It was proven in \cite{Grozdanov:2019kge, Grozdanov:2019uhi} that the radii of convergence of
the hydrodynamic gradient series expansions  are determined by the critical points of the associated complex spectral curves. For
theories with a dual gravitational description through holography, this critical points correspond to level-crossings in the quasinormal spectrum of the dual black hole. 

In this article, we study both the modulated instabilities and physics near the quantum critical AdS$_2$ point in a holographic bottom-up model which has been constructed by a careful comparison with QCD data,  the V-QCD model~\cite{Jarvinen:2011qe}. To this end, we carry out a comprehensive fluctuation analysis of the black hole solutions in this model. We identify and analyze the QNMs in hot and dense holographic QCD plasma, focusing in sectors where most complex and interesting phenomena are expected to arise. Our work significantly extends earlier analysis, where basic transport coefficients (viscosities and conductivities) were computed in hot and dense V-QCD~\cite{Hoyos:2020hmq, Hoyos:2021njg}, and  fluctuations at zero density were studied in the same class of models~\cite{Gursoy:2013zxa,Ishii:2015gia,Demircik:2016nhr,Alho:2020gwl}. This work is also an additional step in the wider program (see, e.g.,~\cite{Alho:2013hsa,Jokela:2020piw,Demircik:2021zll,Demircik:2023lsn,CruzRojas:2023ugm,CruzRojas:2024etx,Demircik:2024bxd}) which aims at establishing and improving predictions for the equation of state and transport properties for dense QCD, based on this model. These predictions can then be used to analyze heavy-ion collisions and neutron star mergers~\cite{Ecker:2019xrw,Tootle:2022pvd,Ecker:2024kzs,Demircik:2022uol}.

The rest of the article is organized as follows. In the rest of the introduction, we give a summary of our results and discuss future directions. In Sec.~\ref{VQCD}, we briefly review the holographic model. We go on discussing the fluctuations in Sec.~\ref{flucs}, including their classification and the derivation of the fluctuation equations. In Sec.~\ref{qnm0q}, we show our results for the quasinormal modes at zero momentum, and derive analytic results for the modes related to the AdS$_2$ geometry. These results are generalized to finite momentum in Sec.~\ref{qnmsfiniteq}, where we focus on the modulated instability and discuss relation to the breaking of hydrodynamics near AdS$_2$ points. The Appendix contains technical details of our analysis.

\subsection{Summary and outlook}

In this article, we carry out a systematic fluctuation analysis of the chirally symmetric charged black hole solutions in V-QCD. We classify all fluctuations, including modes both at nonzero frequency and nonzero momentum. We solve the QNMs found by solving the fluctuation equations numerically, focusing on the most complex sectors of the theory. 
The main results of the article are the following:
\begin{itemize}
    \item We write down the complete set of fluctuation equations for the model in terms of gauge-invariant fluctuation wave functions.
    \item We analyze the fluctuations at the AdS$_2$ IR point, which realizes the quantum critical line of the dual theory at zero temperature and finite density. We compute the dimensions of the fluctuations in the corresponding one-dimensional IR CFT, and showed how the (purely imaginary) QNMs of the black hole phase map to these AdS$_2$ modes as the temperature approaches zero. In particular, we point out an ``anomalous'' behavior of one of the modes, related to the fluctuations of the dilaton field.   
    \item We study the possibility of a modulated instability of gauge fields at finite density, induced by the CS terms which implement the flavor anomaly structure of QCD in the holographic model. We find that the CS terms are strong enough to drive an instability at low temperatures and high densities, as expected from earlier studies~\cite{Domokos:2007kt,Nakamura:2009tf,Ooguri:2010kt,Ooguri:2010xs}. The instability is also present for ``non-Abelian'' fluctuations which transform nontrivially under the flavor group, and the non-Abelian instability is slightly stronger than the Abelian (flavor singlet) instability.
    \item We point out that the region of unstable black holes extends to surprisingly low densities and high temperatures. This finding will be further analyzed in the companion article~\cite{Demircik:2023xxx}.
    \item We analyze the interplay of the hydrodynamic modes with the AdS$_2$ modes. We find similar structure as in~\cite{Arean:2020eus} at low temperature: the convergence of the hydrodynamic dispersion relations for diffusion modes is determined by the collision of the hydrodynamic mode with the leading AdS$_2$ mode. However in our case this happens in a phase which suffers from a modulated instability.
\end{itemize}

This article represents a step towards a realistic holographic model of various ``exotic'' phases that are expected to appear in QCD at high densities. Here we showed that the standard quark matter description is indeed unstable towards an inhomogeneous ground state, in a holographic setup that has been carefully compared to QCD data, in particular to lattice data for thermodynamics. 

There are several future directions to explore:
\begin{itemize}
    \item An obvious task would be the analysis and construction of the end point of the instability, i.e., the modulated ground state. While the simplest possibility is a striped state, where translation invariance is only broken in one direction, more complicated configurations are possible~\cite{Withers:2014sja}. One may try to numerically construct the ground state, or study its nature by carrying out higher order nonlinear fluctuation analysis. This construction is however complicated by the fact that the ground state is likely to be inherently non-Abelian: we found here the instability to be at its strongest in the non-Abelian sector. Notice that the DBI action which we used here is not known for general non-Abelian field configurations, so one needs to first choose a prescription to pin down the action, and the results may depend on the prescription. 
    \item Another task is to figure out how model dependent our predictions for the range of the instability are. This will be discussed in part in the companion article~\cite{Demircik:2023xxx}. This is important as the computation here suggests that instability extends down to low density and high temperature, even to the region which is directly accessible on the lattice.
    \item Apart from the instability towards a modulated state, there might be other kinds of instabilities, in particular those related to various paired phases involving color superconductivity. Such phases are known to be tricky in holography as color superconductivity requires breaking of the gauge symmetry, while the holographic method is normally restricted to color singlets. First attempt could be to construct a simple phenomenological model following the ideas of~\cite{BitaghsirFadafan:2018iqr,Nam:2021ufk}, for example (extended to higher order corrections in the background in \cite{Fadafan:2021xcw} and to the imaginary chemical potential region in \cite{Ghoroku:2020fkv}). Inspiration can also be obtained from top-down approaches to exotic phases, see, e.g.~\cite{Faedo:2017aoe,Faedo:2018fjw,Faedo:2019jlp,Henriksson:2019zph,Henriksson:2021zei}. The possible color superconducting instability would compete with the modulated instability, possibly leading to a rich phase diagram.
    \item An additional important task, which may be related to the extent of the instability, is to develop the flavor dependence of the model and investigate its effect on the modulated and possibly on color superconducting instabilities. This would mean to study the effects of isospin breaking, analyze the symmetry energy in quark matter, and include the strange quark mass in the model. 
    \item A different direction would be to extend the current analysis to include the residues of the QNMs, which might be particularly interesting near the AdS$_2$ region where interesting patterns are found, also to compute numerically the full two-point correlators. In the regime of dense matter, the two point correlators are relevant, for example to analyze neutrino transport~\cite{Jarvinen:2023xrx}.
\end{itemize}
We wish to study all these issues in future work.

\section{The V-QCD model}\label{VQCD}

The V-QCD model~\cite{Jarvinen:2011qe} is a holographic bottom-up model for QCD obtained by combining two building blocks: improved holographic QCD (IHQCD)~\cite{Gursoy:2007cb,Gursoy:2007er}, which models the gluon sector, and a setup for adding flavors through tachyonic Dirac-Born-Infeld (TDBI) actions and the relevant CS terms~\cite{Bigazzi:2005md,Casero:2007ae}. This model is inspired by five-dimensional noncritical string theory but does not have a precise top-down derivation: when the expressions from string theory  fail to produce desired phenomenology, we will switch to a bottom-up approach by modifying and generalizing the action, aiming at a form that can be fitted to QCD data. We will go though the basic contents of the model briefly here; for a thorough discussion, see the review~\cite{Jarvinen:2021jbd}. 

We also remark that while we are using the V-QCD model in this article, many results are not sensitive to the details of the action, and are expected to hold at qualitative level for a larger class of models. In particular, we shall discuss below the physics near the quantum critical AdS$_2$ fixed point which shows some universal features.

\subsection{Action and EoMs}

We wish to model a slightly generalized version of QCD, i.e., a $SU(N_c)$ gauge theory with $N_c$ colors, and $N_f$ quark flavors in the fundamental representation of the gauge group. 
The full field content of the model in the bulk contains duals to the most important, relevant and marginal operators of the theory~\cite{Gursoy:2007cb,Gursoy:2007er,Jarvinen:2011qe}:
\begin{itemize}
    \item The dilaton $\phi$ is dual to the $F^2$ operator, where $F$ is the gluon field in QCD. The corresponding source is the coupling of QCD.
    \item The (bulk) metric is dual to the energy-momentum tensor $T_{\mu\nu}$ of QCD. The source is the metric of the field theory.
    \item The axion $a$ is dual to $F \wedge F$, and the source is the $\theta$-angle.
    \item The tachyon $T^{ij}$, where $i,j=1\ldots N_f$ are the flavor indices, is dual to the quark bilinear $\bar \psi^i \psi^j$. The source is therefore the quark mass matrix.
    \item The left and right handed gauge fields $A_{L/R\, \mu}^{ij}$ are dual to the chiral currents $\bar \psi^i(1\pm \gamma_5)\gamma_\mu\psi^j$. The sources are various chemical potentials (for temporal components) and external fields (for spatial components).
\end{itemize}
Here the first three (last two) fields arise from  the IHQCD (TDBI) sector. 

The action of the model separates to four terms,
\begin{equation} \label{fullaction}
S= S_g + S_f + S_a + S_\mathrm{CS} \ ,
\end{equation}
which describe the gluon sector, the flavor sector, the CP-odd sector linked to the axial anomaly, and a CS sector linked to flavor anomalies in QCD, respectively.
The glue action is given by five dimensional dilaton gravity:
\begin{equation} \label{eq:IHQCDS}
S_g=M_p^3 N_c^2 \int d^{5} x\sqrt{-g}\left[R-\frac{4}{3} g^{M N} \partial_{M} \phi \partial_{N} \phi+V_{g}(\phi)\right] \ ,
\end{equation}
where $M_p$ is the five-dimensional Planck mass.
With an appropriate choice of the dilaton potential $V_g$ this part of the action is that of the IHQCD model, except for the axion field which is included in $S_a$ and is discussed below.

The flavor action $S_f$ is the generalized TDBI action~\cite{Casero:2007ae,Iatrakis:2010jb,Iatrakis:2010zf},

\be
S_f= - \frac{1}{2} M_p^3 N_c\ \mathrm{Tr} \int d^4x\, dr\,
\left(V_f(\phi,T^\dagger T)\sqrt{-\det {\bf A}_L}+V_f(\phi, TT^\dagger)\sqrt{-\det {\bf A}_R}\right)\ .
\label{generalact}
\ee
Here the trace $\mathrm{Tr}$ is over flavor indices, the determinants are over Lorenz indices, and we have denoted
\begin{align}
{\bf A}_{L\,MN} &=g_{MN} + w(\phi) F^{(L)}_{MN}
+ \frac{\kappa(\phi)}{2 } \left[(D_M T)^\dagger (D_N T)+
(D_N T)^\dagger (D_M T)\right] \ ,\nonumber\\
{\bf A}_{R\,MN} &=g_{MN} + w(\phi) F^{(R)}_{MN}
+ \frac{\kappa(\phi)}{2} \left[(D_M T) (D_N T)^\dagger+
(D_N T) (D_M T)^\dagger\right] \  ,
\label{Senaction}
\end{align}
where the covariant derivative is defined as
\be
D_M T = \partial_M T + i  T A_M^L- i A_M^R T\ .
\ee
Recall that the fields  $A_{L}$, $A_{R}$ and $T$ are $N_f \times N_f$ matrices in flavor space. We also denote
 \be
 x\equiv \frac{N_f}{N_c}\ .
 \ee
The string theory inspiration of the model is obtained by considering the Veneziano limit~\cite{Veneziano:1976wm}, where one takes $N_c$ and $N_f$ to infinity keeping $x$ fixed~\cite{Jarvinen:2011qe}. In this limit the flavor sector~\eqref{generalact} is fully backreacted to the glue sector~\eqref{eq:IHQCDS}.
The choice of the potentials $V_f$, $\kappa$, and $w$ will be discussed below.

The determinants over the Lorentz indices in~\eqref{generalact} are ambiguous when the elements contain non-Abelian matrices. For the purposes of this article, however, a precise description is not necessary. All fields of the background solution, which we will soon discuss, will be proportional to the unit matrix $\mathbb{I}_{N_f}$. In particular, we will assume that the quarks are all massless. Therefore non-Abelian fields will only appear when we fluctuate the action around the background solution. We will consider the fluctuated action up to quadratic order, in which case the ambiguity in the ordering of the non-Abelian elements is absent due to the cyclicity of the trace. 

We focus on the chirally symmetric phase (also implying that the quark mass is zero) where the background value of the tachyon vanishes. Then, because the action is quadratic in the tachyon, the tachyon fluctuations will decouple from the rest and can be treated separately. All such quadratic contributions will arise from the DBI action given above. Therefore we can effectively simply set the tachyon to zero in the additional terms of the action that we discuss next. Setting the tachyon to zero, the flavor action simplifies to
\be\begin{aligned}
S_{f0}&= -  \frac{1}{2}M_p^3 N_c \int d^5x\,V_{f0}(\phi)\times\\
&\qquad\times 
\mathrm{Tr}\left(\sqrt{-\det (g_{MN} + w(\phi) F^{(L)}_{MN})}+\sqrt{-\det (g_{MN} + w(\phi) F^{(R)}_{MN})}\right)\ 
\label{generalact2}
\end{aligned}\ee
where $V_{f0}(\phi)=V_f(\phi,T=0)$.

The axion action $S_a$ arises as a combination of terms from the closed and open string sectors~\cite{Arean:2013tja,Arean:2016hcs}. At zero tachyon, it reads
\be \label{Sadef}
 S_a = - \frac{M_p^3}{2} \int d^5x \sqrt{-\det g}\, Z(\phi) \left[N_c \partial_M \hat a - 
 \mathrm{Tr}\left(A^L_{M}-A^R_{M}\right) \right]^2
\ee
where $\hat a = a/N_c$. The choice for the potential $Z(\phi)$ will be analyzed below.  This term is responsible for the holographic implementation of the axial $U(1)_A$ anomaly, and the relative normalization between the gauge field and axion is determined by the anomaly. 

When the tachyon vanishes, the (standard) CS term can be written as~\cite{Witten:1998qj,Casero:2007ae}
\begin{align} \label{eq:SCSdef}
 S_\mathrm{CS} &= \frac{iN_c}{24\pi^2} \int \mathrm{Tr}\bigg[-iA_L\wedge F_L \wedge F_L+\frac{1}{2}A_L\wedge A_L\wedge A_L\wedge F_L + &\nonumber\\ 
 &+ \frac{i}{10}A_L\wedge A_L\wedge A_L\wedge A_L\wedge A_L +iA_R\wedge F_R \wedge F_R-\frac{1}{2}A_R\wedge A_R\wedge A_R\wedge F_R  - &\nonumber\\ 
 &- \frac{i}{10}A_R\wedge A_R\wedge A_R\wedge A_R\wedge A_R \bigg] \ .&
\end{align}
Our sign conventions are such that $F_{L/R} = dA_{L/R} - i A_{L/R}\wedge A_{L/R}$.
The tachyon dependence of the CS term has been analyzed recently in~\cite{Jarvinen:2022mys}.

\subsection{Background}

We then briefly discuss the background we are using in the fluctuation analysis (see~\cite{Alho:2012mh,Alho:2013hsa}).
We set the quark masses and the $\theta$-angle to zero, and consider the chirally symmetric phase. Then the tachyon and axion vanish identically.
The only nonzero component of the gauge fields in the background is the temporal component of the vectorial field, $V_t(r) = \Phi(r) $. We follow a normalization convention where
\be
 A_L = V + A \ , \qquad A_R = V-A \ .
\ee
The flavor Lagrangian for the background is therefore
\be \label{lagrangian}
 \mathcal{L}_{f0} =- M_p^3 N_c N_f V_{f0}(\phi) \sqrt{-\operatorname{det}\left[g_{MN}+w(\phi) F_{MN}\right]}
\ee
where $F_{MN}$ is the field strength tensor for the vectorial gauge field.

The metric Ansatz is
\begin{equation} \label{metric}
\mathrm{d} s^{2} 
=e^{2A(r)}\left[-f(r) \mathrm{d} t^{2}+\mathrm{d} \mathbf{x}^{2}+\frac{\mathrm{d} r^{2}}{f(r)}\right], \quad f(0)=1.
\end{equation}
The nonzero functions describing the background are then the fields $A$ and $f$ in the metric, the dilaton $\phi$, and the gauge field $\Phi$, which all are assumed to depend only on the holographic coordinate $r$. We will only be studying black hole solutions in this article so that $r$ runs from zero at the boundary to a finite value $r_h$ at the black hole horizon, where $f(r_h)=0$. 

With the above assumptions, the background action only depends on $\Phi$ through its $r$-derivative. Therefore the equation of motion for $\Phi$ implies that we have a constant of integration $\hat n$~\cite{Alho:2013hsa}:
\begin{equation} \label{nhatdef}
\frac{1}{M_p^3N_cN_f}\frac{\partial \mathcal{L}_{f}}{\partial \Phi'}=-\frac{e^{A} V_{f0}(\phi) w(\phi)^{2} \Phi'}{\sqrt{1 -e^{-4A}w(\phi)^2 (\Phi')^2 }}\equiv  \hat n = \frac{1}{M_p^3N_cN_f} n \ ,
\end{equation}
where $n$ is the number density of quarks.
From this we can solve to find $\Phi'$ and integrate to determine the quark number chemical potential $\mu$: 
\be
 \mu = -\int_0^{r_h}dr\,\Phi'(r) =\int_0^{r_h}dr \frac{\hat n}{e^{A}V_{f0}(\phi)w(\phi)^2\sqrt{1+\frac{\hat n^2}{e^{6A}w(\phi)^2 V_{f0}(\phi)^2}}}  \ .
\ee
It is important that the weak coupling behavior of the various fields agrees with QCD in order to have the best possible “boundary conditions” for the strong coupling behavior of the model.

We then discuss the background in the weak-coupling region, near the ultraviolet (UV) boundary.
Usually in holographic bottom-up models of QCD one chooses the leading behavior of the bulk fields near the boundary such that it  agrees with the leading UV dimensions of the dual operators. Here we also require that the first few quantum corrections in field theory, and the renormalization group (RG) flow imposed by them, agrees with the near-boundary holographic RG flow of the bulk fields. We demonstrate this here for the coupling~\cite{Gursoy:2007cb}, but the flow of the quark mass and the chiral condensate can be analyzed similarly~\cite{Jarvinen:2011qe}.

The boundary conditions for the tachyon are such
that it vanishes near the boundary. It
also turns out that the gauge field is irrelevant for the
boundary behavior of the metric~\cite{Alho:2013hsa}. Setting $\tau = 0$ in the
action \eqref{fullaction} we see that the geometry is determined by the effective potential 
\be \label{eq:Veffsimple}
V_{\mathrm{eff}}(\l)=V_g(\log\l)-xV_{f0}(\log\l)
\ee
where $\l=e^\phi$ is identified as the 't Hooft coupling near the boundary. For the geometry to be asymptotically AdS$_5$ at the
boundary, $V_\mathrm{eff}$ needs to go to a constant at small coupling:
\begin{equation}
V_{\mathrm{eff}}(\lambda) \rightarrow \frac{12}{\ell^2} \quad \text { as } \quad \lambda \rightarrow 0
\end{equation}
where $\ell$ is the $\mathrm{AdS}_5$ radius for the UV geometry. It is natural to assume a Taylor series around $\lambda=0$, i.e.,
\be
V_{\mathrm{eff}}(\lambda)=\frac{12}{\ell^2}\left[1+v_1 \frac{\lambda}{\lambda_0}+v_2\left(\frac{\lambda}{\lambda_0}\right)^2+\mathcal{O}\left(\lambda^3\right)\right]
\ee
where the constant $\lambda_0=8 \pi^2$ was introduced for  convenience. Then the near boundary asymptotics of the geometry is $\mathrm{AdS}_5$ with logarithmic corrections :
\begin{align}
 A(r)&=-\log \frac{r}{\ell_0}+\frac{4}{9 \log (r \Lambda)} \\ & \quad+\frac{\left(\frac{95}{162}-\frac{32 v_2}{81 v_1^2}\right)+\left(-\frac{23}{81}+\frac{64 v_2}{81 v_1^2}\right) \log (-\log (r \Lambda))}{(\log (r \Lambda))^2}+\mathcal{O}\left(\frac{1}{(\log (r \Lambda))^3}\right)\ , \nonumber \\
\frac{v_1 \lambda(r)}{\lambda_0}&=-\frac{8}{9 \log (r \Lambda)}+\frac{\left(\frac{46}{81}-\frac{128 v_2}{81 v_1^2}\right) \log (-\log (r \Lambda))}{(\log (r \Lambda))^2}+\mathcal{O}\left(\frac{1}{(\log (r \Lambda))^3}\right)\ . &
\label{eq:lambdaUV}
\end{align}
The flow behaves as expected: the source term of the dilaton has become logarithmically flowing instead of a constant, and the value of the source is now identified with the scale $\Lambda = \Lambda_\mathrm{UV}$. This scale defines the units for all dimensionful quantities in our analysis. The field $A(r)$ can be interpreted as the logarithm of the energy scale in field theory. This given a mapping between the series coefficients $v_i$ of the effective potential and the $\beta$-function in QCD~\cite{Gursoy:2007cb,Gursoy:2007er}.

\subsection{Choice of potentials} \label{sec:pot_choice}

The action (\ref{fullaction}) is parametrized in terms of the potentials $V_{g}(\phi), V_{f}(\phi, \tau), \kappa(\phi)$, $w(\phi)$, and $Z(\phi)$ which are chosen to satisfy two basic requirements. First, in the ultraviolet, the model should reproduce the known perturbative behaviors of the corresponding field theory, as we demonstrated for the coupling of QCD above. Second, in the deep infrared the model should lead to the generation of a dynamical wall shielding the singular behavior as $\phi \rightarrow \infty$, which is responsible for confinement in the absence of the tachyon. The details IR wall should be such that it produces among other things, correct behavior of meson masses at large excitation numbers and at finite quark mass, as well as correct kind of phase diagram at finite density. These requirements set tight constraints to the asymptotics of the potentials both at small coupling ($\phi \to -\infty$)~\cite{Gursoy:2007cb,Jarvinen:2011qe} and at large coupling  ($\phi \to \infty$)~ \cite{Gursoy:2007er,Jarvinen:2011qe,Arean:2013tja,Arean:2016hcs,Jarvinen:2015ofa,Ishii:2019gta}.

After the asymptotics of the potentials has been determined, their behavior at intermediate values of $\phi$ needs to be fixed by comparing to QCD data. It is convenient to first fit the functions $V_g(\phi)$ and $Z(\phi)$ in the IHQCD sector to lattice data on the thermodynamics, glueball spectrum, and topological susceptibility of the Yang-Mills theory~\cite{Gursoy:2009jd,Gursoy:2012bt,Jokela:2018ers}. The most challenging step is then to fit the functions $V_{f}(\phi, \tau), \kappa(\phi)$, and $w(\phi)$ to data for full QCD. This has been carried out by using different strategies: by fitting only the lattice data for QCD thermodynamics~\cite{Jokela:2018ers}, by fitting the meson spectrum (including highly excited states)~\cite{Amorim:2021gat}, and recently by fitting both the thermodynamics and spectrum data~\cite{Jarvinen:2022gcc}. In this article, we choose to use the first fit in~\cite{Jokela:2018ers} (in the slightly modified form published in~\cite{Ishii:2019gta}). This fit reproduces the lattice data for the QCD thermodynamics in the deconfined phase to the highest precision, and in this article we will be focusing in this phase.

In~\cite{Jokela:2018ers}, several sets of potentials were determined. We will be using the sets 5b, 7a, and 8b in this article. The different choices reflect the freedom in the parameterization of the potentials which cannot be determined through fitting to lattice data. Most of our analysis will use the intermediate choice, 7a.

Apart from the potentials fitted to thermodynamics in~\cite{Jokela:2018ers,Ishii:2019gta}, we need to specify the function $Z(\phi)$ appearing in the axion term~\eqref{Sadef}. We use an Ansatz introduced in~\cite{Gursoy:2012bt},
\be
 Z(\phi) = Z_0 \left[1 + c_a\, \frac{e^\phi}{8\pi^2} + d_a \left(\frac{e^\phi}{8\pi^2}\right)^4 \right] = Z_0 \left[1 + c_a\, \frac{\l}{\l_0} + d_a \left(\frac{\l}{\l_0}\right)^4 \right] \ .
\ee
Here $Z_0$, $c_a$, and $d_a$ can be fitted to data (see~\cite{Gursoy:2009jd,Gursoy:2012bt}). However, it was noticed in~\cite{Weitz} that a too large value of $Z_0$ can cause a trouble with the backgrounds at finite axial chemical potentials: regular solutions asymptoting to the AdS$_5$ geometry at the boundary no longer exists. Since in this work the axial chemical potential is zero and only the vectorial potential is finite, this problem does not appear, but we want anyhow to choose the function $Z(\phi)$ such that the model is also consistent when the axial potential is nonzero. For the choice of potentials 7a, this implies that $Z_0 \lesssim 0.9060$. Here we follow~\cite{Weitz} and first set the constants $c_a$ and $d_a$ to such values that a small enough $Z_0$ can be obtained. We then determine the value of $Z_0$ by comparing to the lattice results for the topological susceptibility (see Appendix~\ref{app:Z}). Our choice is given by
\be
 Z_0 = 0.670 \ , \qquad c_a = 3 \ , \qquad d_a = 20 \ .
\ee

\section{Fluctuations}\label{flucs}

In this section, we analyze the fluctuations on top of the V-QCD background at finite density and temperature which was introduced in the previous Sec.~\ref{VQCD}. We classify these fluctuations both when the momentum is zero and nonzero. 
We fluctuate all bulk fields as follows:
\begin{equation}\label{fs}
\begin{aligned}
g_{M N}(r,t,z) &=\bar g_{MN}(r)+e^{-i(\omega t-q z)} \delta g_{MN}(r)\ , \\
A_{L/R\,M}^{ij}(r,t,z) &= \bar A_{L/R\,M}(r)\delta^{ij}+e^{-i(\omega t-q z)} \delta A_{L/R\,M}^{ij}(r)\ , \\
\phi(r,t,z) &=\bar  \phi(r)+e^{-i(\omega t-q z)} \delta \phi(r)\ ,\\
T^{ij}(r,t,z) &=e^{-i(\omega t-q z)} \delta T^{ij}(r)\ ,\\
\hat a(r,t,z) &=e^{-i(\omega t-q z)} \delta \hat a(r)\ .
\end{aligned}
\end{equation}
Here the fluctuations are taken in the plane wave form with both finite frequency $\omega$ and momentum $q$, which we choose to be in the $z$-direction. Following Sec.~\ref{VQCD}, only the metric, dilaton, and temporal part of the Abelian vector gauge field have backgrounds, i.e. $\bar  g_{MN}(r)$, $ \bar A_{L/R,M}(r) = \delta_{M}^t \Phi(r)$ and $\bar \phi(r)$, respectively. For notational simplicity, we will not include the bar in the background solution elsewhere in this article. 
As there is no background for the tachyon field, it will decouple from other fluctuations and all the components of the complex matrix $\delta T^{ij}$ satisfy the same equation. Therefore it is enough to take the fluctuation matrix to be proportional to the unit matrix and consider a single wave function:
\be
\delta T^{ij}(r) = \delta^{ij}\delta \tau(r)\ .
\ee
We switch to the basis of vectorial and axial gauge fields, and decompose the fluctuations as 
\be\begin{aligned}
\delta V^M(r) &= \frac{1}{2}\left[\delta A_L^M(r)+\delta A_R^M(r)\right] = \delta \hat V^M(r) \mathbb{I} +  \delta V^{M\, a}(r) t^a  & \\
\delta A^M(r) &= \frac{1}{2}\left[\delta A_L^M(r)-\delta A_R^M(r)\right] = \delta \hat A^M(r) \mathbb{I} +  \delta A^{M\, a}(r) t^a  & \\
\end{aligned}\ee
where $\mathbb{I}$ represents the $N_f\times N_f$ unit matrix in flavor space, and the $SU(N_f)$ generators are normalized as $\mathrm{Tr}(t^at^b)=\delta^{ab}/2$. 
From now on, we refer to the flavors singlet terms $\delta \hat{V}^M$ and $\delta \hat{A}^M$ as Abelian 
fluctuations, and $\delta V^{M\,a}$ and $\delta A^{M\,a}$ as non-Abelian 
fluctuations, respectively. The division to these components is natural: because the background is flavor singlet, flavor symmetry dictates that fluctuations in the Abelian and non-Abelian sectors are decoupled.

\subsection{Classification of fluctuations and the gauge invariant combinations}\label{givs}

We then discuss how the fluctuations are reduced to the physically relevant components and how they can be classified (see, e.g., \cite{Ammon:2017ded,Jansen:2017oag}). In order to do this, we first recall how the various fields transform under diffeomorphisms and bulk gauge transformations:
\begin{align} 
    g_{MN}&\mapsto g_{MN}-\nabla_{M}\xi_{N}-\nabla_{N}\xi_{M},
\nonumber\\ 
    \hat{V}_{M}&\mapsto\hat{V}_{M} -\partial_{M}\alpha-\xi^{K}\nabla_{K}\hat{V}_{M}-\hat{V}_{K}\nabla_{M}\xi^{K},\nonumber\\
 \hat{A}_{M}&\mapsto\hat{A}_{M}-\partial_{M}\beta-\xi^{K}\nabla_{K}\hat{A}_{M}-\hat{A}_{K}\nabla_{M}\xi^{K},
\nonumber\\ 
 V^{a}_{M}&\mapsto V^{a}_{M}-\partial_{M}\alpha^a-f^{abc}V_M^b\alpha^c-\xi^{K}\nabla_{K}V_{M}^a-V_{K}^a\nabla_{M}\xi^{K},\nonumber\\
    A^{a}_{M}&\mapsto A^{a}_{M}-\partial_{M}\beta^a-f^{abc}A_M^b\beta^c-\xi^{K}\nabla_{K}A_{M}^a-A_{K}^a\nabla_{M}\xi^{K},\nonumber\\
    \phi&\mapsto \phi-\xi^{K}\nabla_{K}\phi,\qquad\
    \tau\mapsto \tau-\xi^{K}\nabla_{K}\tau, \qquad
     \hat a\mapsto \hat a- \frac{2N_f}{N_c} \beta -\xi^{K}\nabla_{K}\hat a.
 \label{eq:gaugeT}
 \end{align}
Here, $\alpha$, $\beta$, are the gauge functions for the Abelian vector and axial vector fields, $\alpha^a$ $\beta^a$ are the gauge functions for the non-Abelian vector and axial fields, and  $\xi_M$ are the gauge functions for the diffeomorphisms (infinitesimal variations of the  coordinates).
In addition, $f^{abc}$ is the structure constant defined via $[t^a,t^b]=if^{abc}t^c$, and the covariant derivatives are 
defined by using the  
background metric 
given in equation~\eqref{metric}.

The total number of degrees of freedom (DOFs) in the fluctuations in the Abelian sector is 28, which includes 15 from the metric, 10 from the gauge fields, and 3 scalars. Fixing the 7 Abelian gauge transformations in~\eqref{eq:gaugeT}, which we do here by imposing the radial gauge  ($\delta g_{Mr}=0$, $\delta \hat{V}_{r}=0$, and $\delta \hat{A}_{r}=0$), the number of DOFs is reduced to 21. These include nondynamical DOFs, which are eliminated by constraint equations (see~\cite{Kiritsis:2006ua} for a detailed discussion of the gravity sector). The number of dynamical DOFs is 14, which includes a four dimensional spin-two graviton (5 DOFs), vector and axial spin-one fields (6 DOFs), and three scalars (3 DOFs).

In the non-Abelian sector the counting of DOFs is somewhat simpler. We ignore the non-Abelian components of the tachyon that we have already left out of the analysis. Then the DOFs are the non-Abelian vector and axial fields, which have in total 10 multiplets in the adjoint representation of the vectorial $SU(N_f)$. After imposing the gauge conditions  $\delta V^a_{r}=0$, $\delta A^b_{r}=0$ and eliminating nondynamical DOFs, we are left with 6 multiplets which are recognized as non-Abelian the vector and axial spin-one fields on the boundary.

\begin{table}[ht!]
\centering
\begin{tabular}{|c || c | c | c | c || c | c |} 
 \hline
  & spin 2  & spin 1  & spin 1 & spin 0 & spin 1 & spin 1 \\ 
  & & Abelian & Abelian & & non-Abelian & non-Abelian  \\ 
  & graviton & vector & axial & & vector & axial \\ 
 \hline  \hline
 helicity $\pm$2 & 2 & 0 & 0 & 0 & 0 & 0 \\ 
 \hline
 helicity $\pm$1 & 2 & 2 & 2 & 0 & 2 & 2  \\
 \hline
 helicity 0 & 1 & 1 & 1 & 3 & 1 & 1  \\
 \hline
\end{tabular}
  \caption{ The number of degrees of freedom classified according to helicity, spin and parity  in Abelian and Non-Abelian sectors. Notice that here we left out the dimension of the adjoint representation of $SU(N_f)$ under which the non-Abelian fluctuations transform. Each multiplet in this representation is counted as one in the table.}
    \label{doftable}
\end{table}

A summary of the dynamical fluctuations is given in Table~\ref{doftable}.
Apart from the properties already discussed above, the fluctuations can be classified according to helicity or equivalently the $z$-component of the spin, i.e., their transformation properties under the  
$SO(2)$ rotation symmetry around the $z$-axis, fixed by the direction of the momentum.
As we have chosen the momentum to lie in the $z$-direction, this rotation is an exact symmetry even at  finite momentum. Apart from the decoupling of Abelian and non-Abelian fluctuations, the fluctuations with different helicities therefore decouple, as the background is rotation invariant. Notice that fluctuations with different spins may be coupled.
Apart from the fields listed in the table, there are in principle also scalar and pseudoscalar non-Abelian fields arising from the fluctuated tachyon, as well as an additional Abelian scalar mode corresponding to the ``imaginary part'' of the tachyon fluctuation. However as we pointed out above, these fields satisfy the same equation as the Abelian tachyon field, and can therefore be excluded without loss of generality.

Even though we have already fixed the gauge, it is still useful to write the fluctuation equations in terms of gauge invariant variables (GIVs), i.e., variables invariant under the transformations~\eqref{eq:gaugeT} for plane wave gauge functions (for example $\alpha(x_M) = e^{-i(\omega t - qz)}\alpha(r)$). This is because if the variables are gauge invariant, it is guaranteed that all coefficients in the fluctuation equations are gauge independent as well, which typically leads to the simplest expressions for these equations. Notice that this means that actually gauge fixing would not be necessary to obtain the final equations, but it simplifies the steps in their derivations as there are less functions to keep track of.

As fluctuations with different helicities are decoupled, we discuss the GIVs for each helicity channel separately. To obtain the fluctuation equations, we insert the Ansatz in~\eqref{fs} to the equations of motion listed in Appendix~\ref{app:eoms}, and express the fluctuations of these equations in terms of the GIVs listed here. The resulting fluctuation equations are given in Appendix~\ref{flucEq}.

\paragraph{Helicity $0$:}
The GIVs in the Abelian sector are 
\begin{equation}
\begin{aligned}    &Z_0=e^{2A(r)}\left[-q^{2} f(r) \delta g_{tt}+2  \omega q \delta g_{tz}\frac{}{}\right.\nonumber\\
&\qquad\qquad\qquad\qquad\left.+  \frac{1}{2}\left(q^2 f(r)-\omega^{2}+\frac{q^2 f'(r)}{2A'(r)}\right)(\delta g_{xx}+\delta g_{yy})+ \omega^2 \delta g_{zz}\right],  \\
&\hat E_V=q\delta \hat V_{t}+\omega \delta \hat V_{z}-\frac{(\delta g_{xx}+\delta g_{yy})q \Phi'}{4 A'(r)}\ ,  \qquad\qquad\;\;
Z_{\phi}=\delta \phi-\frac{\phi^{\prime}(\delta g_{xx}+\delta g_{yy})}{4 A^{\prime}(r)}\ ,  \\
&Z_a = \delta \hat a -2 i x\frac{ \omega  \delta \hat A_t/f+q \delta \hat A_z}{\omega ^2/f-q^2}\ , \qquad\quad \quad\;\qquad\quad\;\;
\hat E_A=q\delta \hat A_{t}+\omega \delta \hat A_{z}\ ,
\label{hel0fldefs}
\end{aligned}
\end{equation}
 and the tachyon fluctuation, $\delta \tau$ above, which is readily diffeomorphism invariant in the absence of a tachyon background.
 The non-Abelian GIVs are
 \be
E_V^a=q\delta V_{t}^a+\omega \delta V_{z}^a\ , \qquad\qquad\qquad\qquad\;\;\;\qquad\quad\;\;
E_A^a=q\delta A_{t}^a+\omega \delta A_{z}^a\ .
\ee

 The abelian sector further decouples into subsectors of $\{Z_0,\hat{E} _V,Z_\phi\}$ (where metric, Abelian vector and dilaton fluctuations couple), $\{Z_a,\hat{E}_A\}$ (where axion, Abelian axial fluctuations couple) and $\{\delta\tau\}$ (the tachyon is decoupled from other fields).  The coupling of the subsectors comprising vector and axial Abelian gauge field fluctuations stems from different signatures under parity transformation. In non-Abelian sector, the GIVs $\{E_V^a\}$ and $\{E_A^a\}$ decouple, again due to parity, moreover they satisfy the same equations.

\paragraph{Helicity $\pm1$:} The 
GIVs in the Abelian sector are 
\begin{equation}
\begin{aligned}
&Z_1^\pm= e^{2A}\left[q \left(\delta g_{tx}\pm i\delta g_{ty}\right)+ \omega \left(\delta g_{zx}\pm i\delta g_{zy}\right)\right]\ ,\\
& \delta \hat V^{\pm} = \delta \hat V_x \pm i\delta \hat V_y\ , \qquad\quad\;  \delta \hat A^{\pm} = \delta \hat A_x \pm i\delta \hat A_y\ ,
\end{aligned}
\end{equation}
and the GIVs in the non-Abelian sector are
\be
 \delta V^{\pm\, a} = \delta V_x^a \pm i\delta V_y^a \ , \qquad  \delta A^{\pm\, a} = \delta A_x^a \pm i\delta A_y^a\ ,
\ee
where the $\pm$ sign denote different helicities. The GIVs with positive and negative helicity are decoupled by rotation symmetry so it is enough
to consider the  
fields with positive helicity.

Abelian sector comprises the set $\{Z_1^\pm,\delta\hat V^\pm,\delta\hat A^\pm\}$. Therefore it involves coupled fluctuations  of the metric, vector and the axial spin-one fields.
In the non-Abelian sector of $\{\delta V^{\pm a},\delta A^{\pm a}\}$, non-Abelian vector and axial fluctuation couple through the CS term. Actually, the structure of the CS term is such that the left and right handed non-Abelian fields are decoupled (see Appendix~\ref{flucEq}).

\paragraph{Helicity $\pm2$:} We choose the 
GIVs to be the helicity eigenstates
\be
Z_2^\pm = e^{2A}\left(\delta g_{xx} - \delta g_{yy}\pm 2 i \delta g_{xy} \right)\ ,
\ee
we note that actually any linear combination of $\delta g_{xx}-\delta g_{yy}$ and $\delta g_{xy}$ works equally well because the fields are decoupled and satisfy the same fluctuation equation.

\begin{figure}[!tb]
\centering
\includegraphics[height=0.42\textwidth]{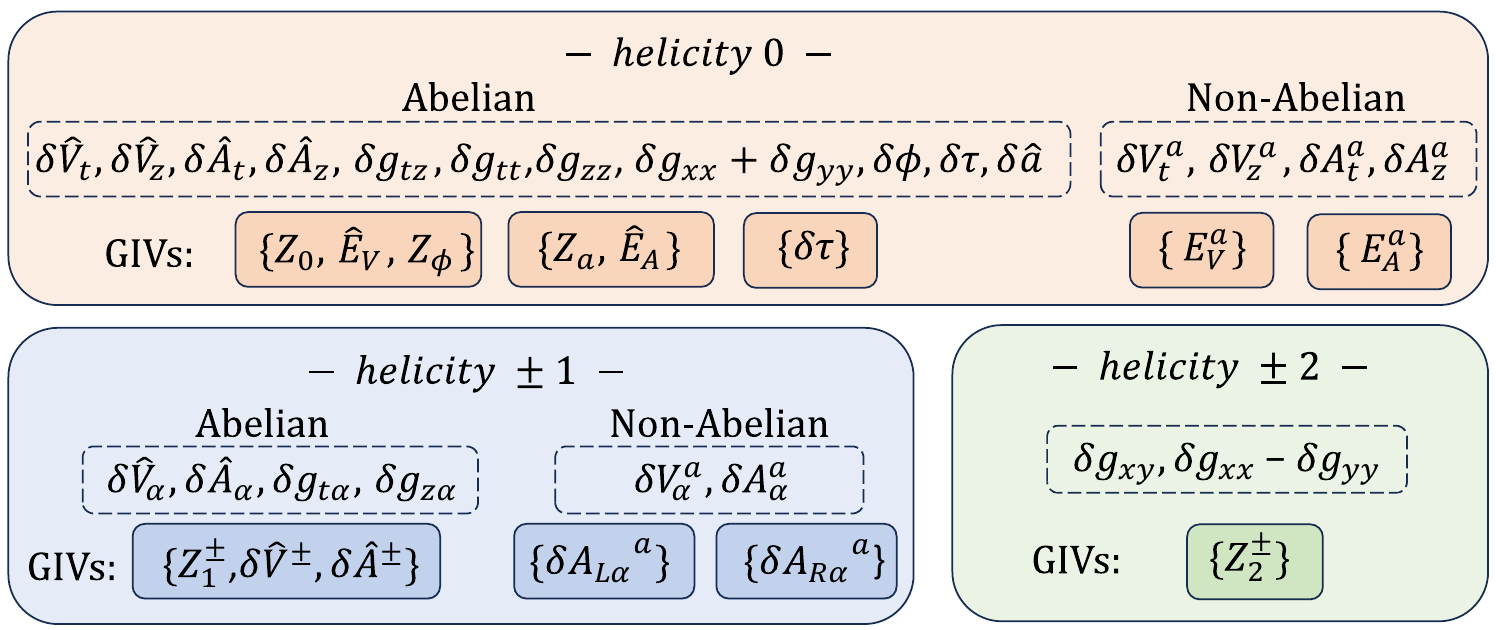}
   \caption{\small Finite momentum classification  
   of fluctuations (represented by boxes with dashed frames) and the coupling among gauge invariant combinations (GIVs, depicted by boxes with solid frames) in the different helicity channels.}
    \label{finq_class}
\end{figure}

\paragraph{Summary:} The classification of the  fluctuations and coupling among GIVs is 
recapitulated 
in Figure~\ref{finq_class}. In the figure, $\alpha$ is used to denote spatial components that are transverse to the momentum direction, i.e. $\alpha=x,y$. The GIVs obey systems of  second order differential equations according to the coupling patterns described above and in Figure~\ref{finq_class}. Explicit expressions for the fluctuation equations are
given 
in Appendix~\ref{flucEq}. 

\subsection{Limit of zero momentum} In the limit of zero momentum there is no preferred spatial direction and the full rotation symmetry of the fluctuations is restored.
This implies that fluctuations with different spin are decoupled, and the fluctuations belonging to the same spin multiplets (given in Table~\ref{doftable})  satisfy the same equations. 
Actually, all the fluctuations are decoupled in this limit.  
Thus the fluctuations are classified into spin $0$, spin $1$ and spin $2$ as follows:

\paragraph{Spin $0$:} This channel consists of the GIVs $Z_\phi$, $Z_a$, $\delta\tau$, corresponding to the three scalars: the dilaton, the axion, and the tachyon, respectively, which each satisfy different equations (see Appendix~\ref{app:q0flucts}). 
\paragraph{Spin $1$:} The Abelian vectorial fields $E_V$ and $\delta\hat V^\pm$ form one spin-one multiplet, and the Abelian axial fields $E_A$ and $\delta\hat A^\pm$ form another.  The vector and the axial fluctuation equations differ in their mass terms. The structure of the non-Abelian fields is similar: the vector  fields ($E_V^a$ and $\delta V^{\pm\,a}$) form one multiplet and the axial fields ($E_A^a$ and $\delta A^{\pm\,a}$) form another. In this case all fields satisfy the same fluctuation equation. 
\paragraph{Spin $2$:} The GIVs $Z_0$ from helicity $0$, $Z_1^\pm$ from helicity $1$ and $Z_2$ from helicity $2$ channels form the graviton multiplet and all obey the same equation.
\paragraph{Summary:} The classification into spin channels and the mapping among GIVs are
recapitulated in the schema depicted in Figure~\ref{q0class}. The decoupled equations obeyed by the GIVs in the limit of vanishing momentum are given in Appendix~\ref{app:q0flucts}.
\begin{figure}[ht!]
\centering
\includegraphics[height=0.128\textwidth]{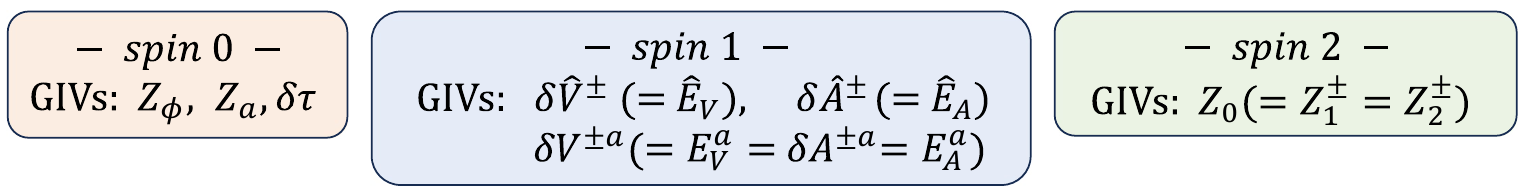}
   \caption{\small Zero momentum classification into different spin channels.  
   The equality sign indicates that the modes satisfy the same fluctuation equations.}
    \label{q0class}
\end{figure}
\section{QNMs at zero momentum}\label{qnm0q}

In this section, we conduct a detailed analysis of QNMs at zero momentum. This limit is special as it results in the decoupling of all fluctuations. In the first subsection,
we study the quantum critical low-temperature and high-density regime, where the geometry is controlled by 
an AdS$_2$ fixed point~\cite{Liu:2009dm,Faulkner:2009wj,Iqbal:2011in}, which was has been analyzed in the V-QCD model in~\cite{Alho:2013hsa,Hoyos:2021njg}.  
We 
determine the locations of the QNMs analytically for the AdS$_2$ geometry and compare these results with the numerical solutions. Subsequently, in the second subsection, we thoroughly explore the temperature dependence of the QNMs at zero momentum. 

\subsection{Behavior at the AdS\texorpdfstring{$_2$}{TEXT} point}\label{ads2}

Let us then analyze the behavior at low temperatures, in the region controlled by the AdS$_2$ point.
Before analyzing the fluctuations, it is useful to remind how the geometry of the solutions in the deconfined phase at finite density, $\mu \sim \Lambda$, changes as the temperature varies:
\begin{itemize}
    \item When $T=0$, the geometry is asymptotically AdS$_5$ near the boundary, and asymptotes to AdS$_2\times \mathbb{R}^3$ in the IR.
    \item When $0<T \ll \Lambda$, the flow is otherwise similar to the zero temperature solution, but the ``pure'' AdS geometry is replaced by a AdS$_2$ black hole. This is a general structure for a near-extremal charged black hole, and it has been analyzed in detail in~\cite{Hoyos:2021njg} in the context of V-QCD.
    \item When $T \sim \Lambda$, the geometry is that of a ``regular'' charged black hole: The geometry is asymptotically AdS$_5$ near the boundary, but does not have other regions where the geometry takes a simple form.
\end{itemize}

The fixed point AdS$_2 \times \mathbb{R}^3$ geometry (without flow) is also an exact solution of the V-QCD action. This solution is found by setting $A$ and $\phi$ to constants, which we denote by $A_*$ and $\phi_*$, respectively. In this case the equations of motion (see~\eqref{feq}--\eqref{dileq} in Appendix~\ref{app:eoms}) imply~\cite{Alho:2013hsa}
\be \label{ads2conds}
 V_\mathrm{eff}(\phi_*,A_*,\hat n) = 0 = \frac{\pa V_\mathrm{eff}(\phi_*,A_*,\hat n)}{\pa \phi} \ ,
\ee
where 
\begin{equation}\label{Veffdeftext}
 V_\mathrm{eff}(\phi,A,\hat n) \equiv V_g(\phi) - x V_{f0}(\phi) \sqrt{1+\frac{\hat n^2}{e^{6A}V_{f0}(\phi)^2w(\phi)^2}} \ .
\end{equation}
These equations determine the value of $\phi_*$ and the combination
\be
 \tilde n_* = \frac{\hat n}{e^{3A_*}}
\ee
which is invariant under the symmetries~\eqref{rtrans} and~\eqref{ftrans} of the equations of motion, given in Appendix~\ref{app:eoms}. This variable is identified as the ratio between the number density and the entropy~\cite{Alho:2013hsa}. Choosing the boundary of the AdS$_2$ space (which is interpreted as the extremal horizon of the full five-dimensional geometry) to lie at $r=0$, the solution for the blackening factor is obtained from~\eqref{feq}:
\be \label{fads2}
 f(r) = \frac{e^{2A_*}}{L_2^2} r^2
\ee
where
\be \label{lads2}
 L_2 =\frac{1}{\sqrt{\frac{1}{6}\frac{\pa V_\mathrm{eff}(\phi_*,A_*,\hat n)}{\pa A}}}
\ee
is the radius of the AdS$_2$ space. The resulting geometry is
\be \label{eq:AdS2metric}
 ds^2 =  \frac{L_2^2dr^2}{r^2} - \frac{e^{4A_*}r^2dt^2}{L_2^2} + e^{2A_*}d\mathbf{x}^2 \ .
\ee
Notice that when this geometry is found as the endpoint of the zero temperature RG flow, the only free parameter is $A_*$ while rest are fixed by the potentials. This parameter is interpreted as (the logarithm of) the IR energy scale~\cite{Alho:2013hsa,Hoyos:2021njg}, and solving its relation to the UV scale $\Lambda$ in~\eqref{eq:lambdaUV} requires the knowledge of the whole flow.  When the chemical potential is varied keeping $T=0$ (and $\Lambda$) fixed, the flow and therefore also the value of $A_*$ change. However, this parameter will cancel in the expressions for the dimensions we derive below.

Let us then study the fluctuations in the AdS$_2$ background. The presence of the AdS factor in the geometry is interpreted to signal the presence of a one dimensional IR CFT~\cite{Faulkner:2009wj,Edalati:2010hk}. Even though we do not have an explicit dictionary between the fluctuations around the fixed point and the operators they correspond to in the CFT, the dimensions of the fluctuations (or the corresponding operators in the CFT) are still useful to classify them.
For this purpose, it is enough to look at the fluctuation equations at zero frequency and momentum (using the general expressions from Appendix~\ref{app:q0flucts}) for the exact AdS$_2 \times \mathbb{R}^3$ geometry. 

Using the expressions from Appendix~\ref{app:q0flucts}, we see that the fluctuations of the metric and non-Abelian gauge fields satisfy the equation
\be
 F''+\frac{2}{r}F' = 0 \ ,
\ee
where $F$ stands for any fluctuation wave functions in the metric and non-Abelian gauge-field sectors.
This equation has the solution
\be
 F
 =C_1 r^{\Delta_*-1} +C_2 r^{-\Delta_*} 
\ee
with $\Delta_*=1$. 

Another set of fluctuations that turns out to obey a simple fluctuation equation in the AdS$_2$ geometry is the set of Abelian vector fluctuations. Their fluctuation equation in~\eqref{Abelq0eqs} has the additional mass term
\be
\frac{\hat n^2 x e^{-4A}}{f V_{f0}(\phi) w(\phi )^2 R} = \frac{1}{3}\frac{e^{2A}}{f}\frac{\pa V_\mathrm{eff}(\phi,A,\hat n)}{\pa A}
\ee
so at the AdS$_2$ point, and at zero frequency, we find (using~\eqref{fads2} and~\eqref{lads2}) that
\be 
 F_V''+\frac{2}{r}F_V'-\frac{2}{r^2}F_V = 0 \ ,
\ee
where $F_V$ stands for all the different helicity components of the Abelian vector fluctuations.
This is the equation for a mode with $\Delta_*=2$. The fluctuations of the metric and the gauge field, except for the Abelian axial field, are therefore ``universal'' modes with integer $\Delta_*$.

The rest of the modes are ``nonuniversal'', i.e., they have noninteger dimensions. First, for the Abelian axial field we find
\be 
 F_A''+\frac{2}{r}F_A'-\frac{8 Z(\phi_*)}{\tilde n_*^2r^2}F_A = 0 \ ,
\ee
from which we read
\be
 \Delta_* = \frac{1}{2} + \frac{1}{2} \sqrt{1 +\frac{32 Z(\phi_*)}{\tilde n_*^2}} \ .
\ee
For the tachyon fluctuation, we find
\be 
 \delta \tau''+\frac{2}{r}\,\delta \tau + \frac{4 w(\phi_*)^2}{x\tilde n_*^2\kappa(\phi_*)\sqrt{1+\frac{\tilde n_*^2}{V_{f0}(\phi_*)^2w(\phi_*)^2}}}\frac{1}{r^2}\,\delta \tau = 0 \ ,
\ee
and 
\be
 \Delta_* = \frac{1}{2} + \frac{1}{2} \sqrt{1 -\frac{16 w(\phi_*)^2}{x\tilde n_*^2\kappa(\phi_*)\sqrt{1+\frac{\tilde n_*^2}{V_{f0}(\phi_*)^2w(\phi_*)^2}}}} \ .
\ee 
Moreover, the axion fluctuation equation becomes
\be 
 Z_a''+\frac{2}{r}Z_a'-\frac{8 Z(\phi_*)}{\tilde n_*^2r^2\left(1+\frac{\tilde n_*^2}{V_{f0}(\phi_*)^2w(\phi_*)^2}\right)}Z_a = 0 \ ,
\ee
from which we read
\be
 \Delta_* = \frac{1}{2} + \frac{1}{2} \sqrt{1 +\frac{32 Z(\phi_*)}{\tilde n_*^2\left(1+\frac{\tilde n_*^2}{V_{f0}(\phi_*)^2w(\phi_*)^2}\right)}} \ .
\ee

The analysis of the dilaton equation~\eqref{Zphiq0eq} however turns out to be a bit involved. Naively imposing the conditions~\eqref{ads2conds} leads to 
\be \label{dilads2eq}
 Z_\phi''+\frac{2}{r}Z_\phi'-\frac{9\, \partial_\phi^2 V_\mathrm{eff}(\phi_*,A_*,\hat n)}{4\,\partial_A V_\mathrm{eff}(\phi_*,A_*,\hat n)}\frac{1}{r^2}Z_\phi = 0 \ ,
\ee
which gives
\be \label{dildeltaads}
 \Delta_* = \frac{1}{2}\left[1+\sqrt{1-\frac{9\, \partial_\phi^2 V_\mathrm{eff}(\phi_*,A_*,\hat n)}{\partial_A V_\mathrm{eff}(\phi_*,A_*,\hat n)}}\right] = \frac{1}{2}\left[1+\sqrt{1-\frac{3}{2} L_2^2 \partial_\phi^2 V_\mathrm{eff}(\phi_*,A_*,\hat n)}\right] \ .
\ee
However, in general the denominators in the mass term of~\eqref{Zphiq0eq} have factors of $A'$ which also vanishes at the AdS$_2$ point. This reflects the fact that $Z_\phi$ in~\eqref{hel0fldefs} becomes ill defined for the AdS$_2$ solution. When the dilaton does not flow, the fluctuation $\delta \phi$ is diffeomorphism invariant, and should be used instead of $Z_\phi$. Working out the fluctuations directly around the fixed AdS$_2 \times \mathbb{R}^3$ geometry of Eq.~\eqref{eq:AdS2metric}, one can check that~\eqref{dilads2eq} is indeed the correct fluctuation equation for the dilaton in this case.
 
The situation for the flows ending on the AdS$_2$ point is however more complicated. The flow solution near the AdS$_2$ point that has been found explicitly in~\cite{Hoyos:2021njg}, and can be inserted in the general fluctuation equation~\eqref{Zphiq0eq}. As it turns out (see Appendix~\ref{app:q0flucts}), the equation~\eqref{dilads2eq} is obtained when $\Delta_*\ge 3/2$ in~\eqref{dildeltaads}, but not when $\Delta_*<3/2$. The general expression for the dimension in this case is:
\begin{align} \label{dildeltaflow}
 \Delta_*^\mathrm{fl} &= \mathrm{max}\left\{\Delta_*,3-\Delta_*\right\}&\\\nonumber
 &= \mathrm{max}\left\{\frac{1}{2}\left[1+\sqrt{1-\frac{9\, \partial_\phi^2 V_\mathrm{eff}(\phi_*,A_*,\hat n)}{\partial_A V_\mathrm{eff}(\phi_*,A_*,\hat n)}}\right],\frac{1}{2}\left[5-\sqrt{1-\frac{9\, \partial_\phi^2 V_\mathrm{eff}(\phi_*,A_*,\hat n)}{\partial_A V_\mathrm{eff}(\phi_*,A_*,\hat n)}}\right]\right\} \ . &
\end{align}

\begin{table}[]
    \centering
    \begin{tabular}{|c|c|c|c|c|}
    \hline
        Field & Dimension $\Delta_*$ & 5b & 7a & 8b  \\
        \hline 
        \hline 
        $\phi$ & $\frac{1}{2}\left[1+\sqrt{1-\frac{3}{2} L_2^2 \partial_\phi^2 V_\mathrm{eff}(\phi_*,A_*,\hat n)}\right]$ & 1.3558 & 1.6566 & 1.4850  \\ 
          & $\Delta_*^\mathrm{fl} = \mathrm{max}\left\{\Delta_*,3-\Delta_*\right\}$ & 1.6442 & 1.6566 & 1.5150 \\ 
        \hline 
        $\tau$ &$\frac{1}{2} + \frac{1}{2} \sqrt{1 -\frac{16 w(\phi_*)^2}{x\tilde n_*^2\kappa(\phi_*)\sqrt{1+\frac{\tilde n_*^2}{V_{f0}(\phi_*)^2w(\phi_*)^2}}}}$ & 0.9967 & 0.9946 & 0.9905  \\ 
        \hline 
        $a$ & $\frac{1}{2} + \frac{1}{2} \sqrt{1 +\frac{32 Z(\phi_*)}{\tilde n_*^2\left(1+\frac{\tilde n_*^2}{V_{f0}(\phi_*)^2w(\phi_*)^2}\right)}}$ & 1.0008 &  1.0034 & 1.0180 \\ 
        \hline 
        $\hat A_\mu$ & $\frac{1}{2} + \frac{1}{2} \sqrt{1 +\frac{32 Z(\phi_*)}{\tilde n_*^2}}$ & 1.0266 & 1.0382 & 1.0606  \\ \hline 
    \end{tabular}
    \caption{The nonuniversal dimensions at the AdS$_2$ fixed point. We include both the analytic expressions and the numerical values for the potential sets used in this article. 
    Fluctuated fields from top to bottom rows are the dilaton, the tachyon, the axion, and the Abelian axial gauge field.}
    \label{tab:dimensions}
\end{table}

The results for the nonuniversal modes are collected in Table~\ref{tab:dimensions}. 
Note that apart from the dimension of the dilaton, the anomalous dimensions are small: the dimensions at the fixed point deviate from one by at most a couple of per cent. 
The value of the charge $\tilde n_*$ at the fixed point, which is controlled by the fit to the lattice data, turns out to be large as compared to the values of the potentials. Typically we find that $\tilde n_* = \mathcal{O}(10)$, while $w(\phi)=\mathcal{O}(1)$, $\kappa(\phi)=\mathcal{O}(1)$ and $Z(\phi)=\mathcal{O}(1)$. The value of $V_{f0}(\phi_*)$ might be a bit larger than other potentials, but still less than $\hat n_*$. 
Because in $n_*$ appears in the denominators in the terms in the square root expressions that cause the dimensions of Table~\ref{tab:dimensions} to deviate from one, the deviations are small.

Let us then recall what these results for the AdS$_2$ dimensions mean for the QNMs. At nonzero but small temperature, where the geometry approaches the AdS$_2$ form in the IR, we expect that a sequence of ``AdS$_2$ modes'' is found in each sector, with frequencies given by the formula~\cite{Faulkner:2010tq,Arean:2020eus}
\be\label{Ads2qnm}
 \frac{\omega}{2\pi T} = -i\left(\Delta_* + n\right) 
\ee
where $n=0,$ $1$, $2$, \ldots. 
We will verify this explicitly below for the $n=0$ modes. In particular, we find that the dilaton QNMs converge to the value given by~\eqref{dildeltaflow} when it is distinct from~\eqref{dildeltaads}.

\begin{figure}[!tb]
\centering
\includegraphics[width=12cm]{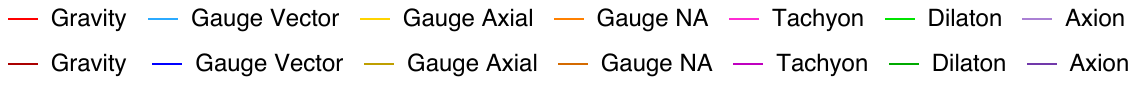}
 \includegraphics[width=7.3cm]{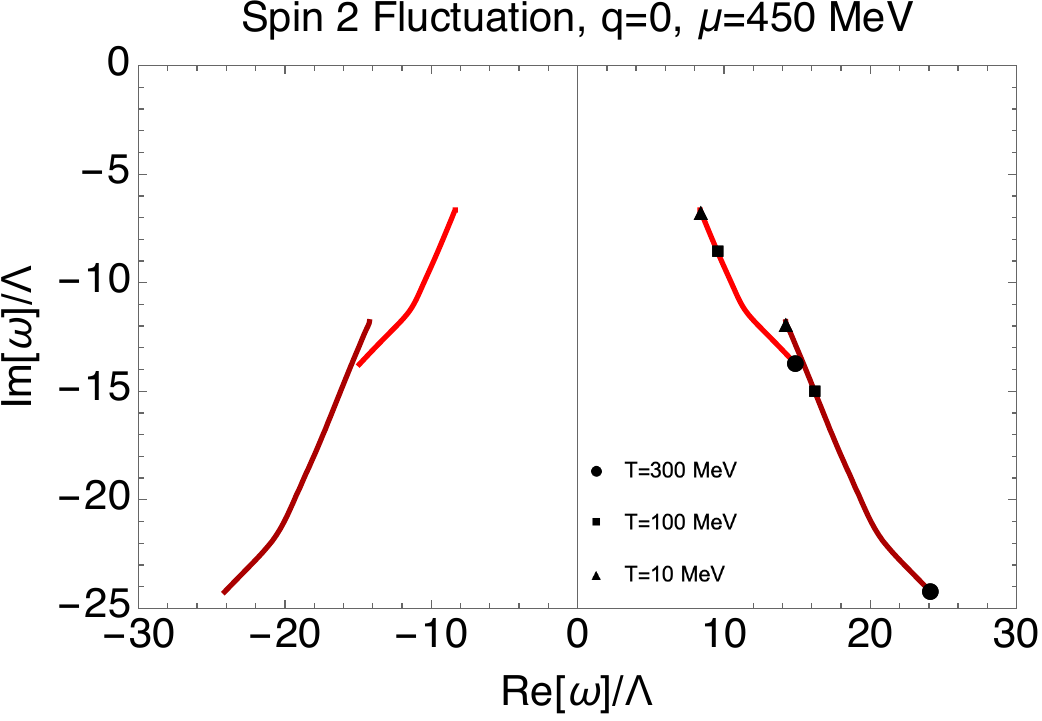}%
 \includegraphics[width=7.3cm]{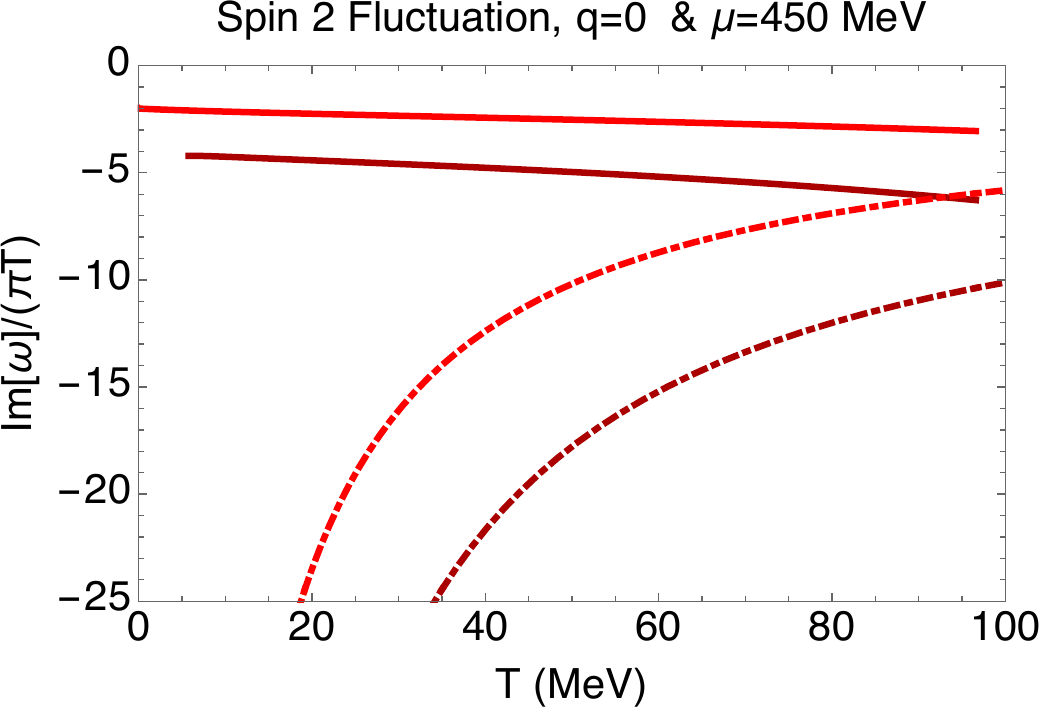}    \includegraphics[width=7.3cm]{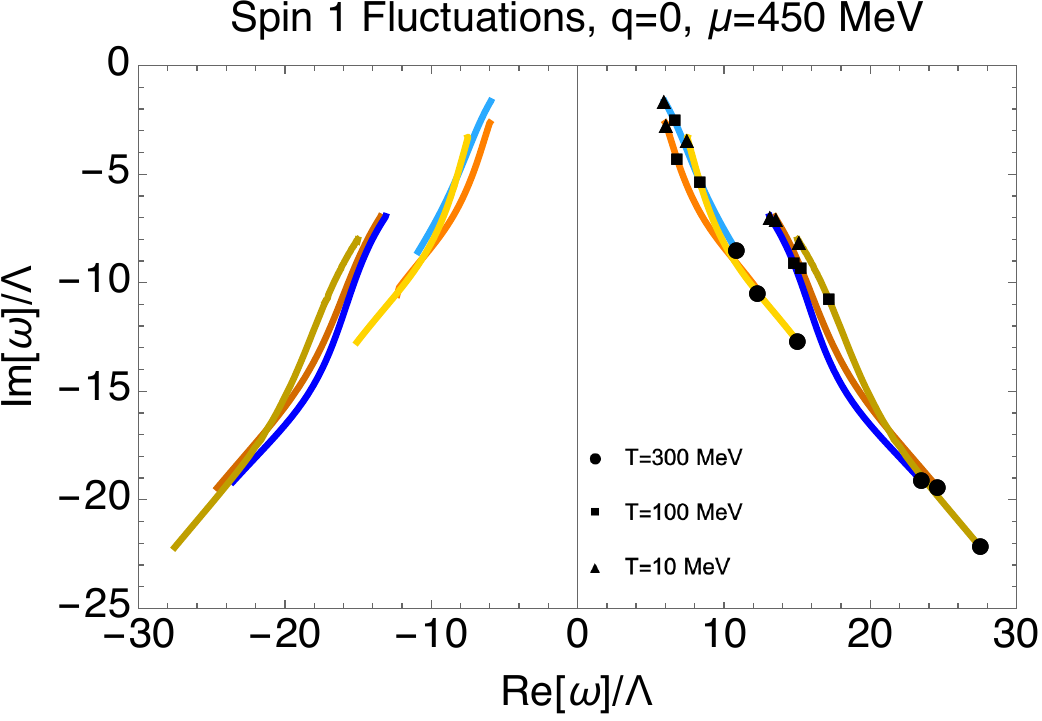}%
 \includegraphics[width=7.3cm]{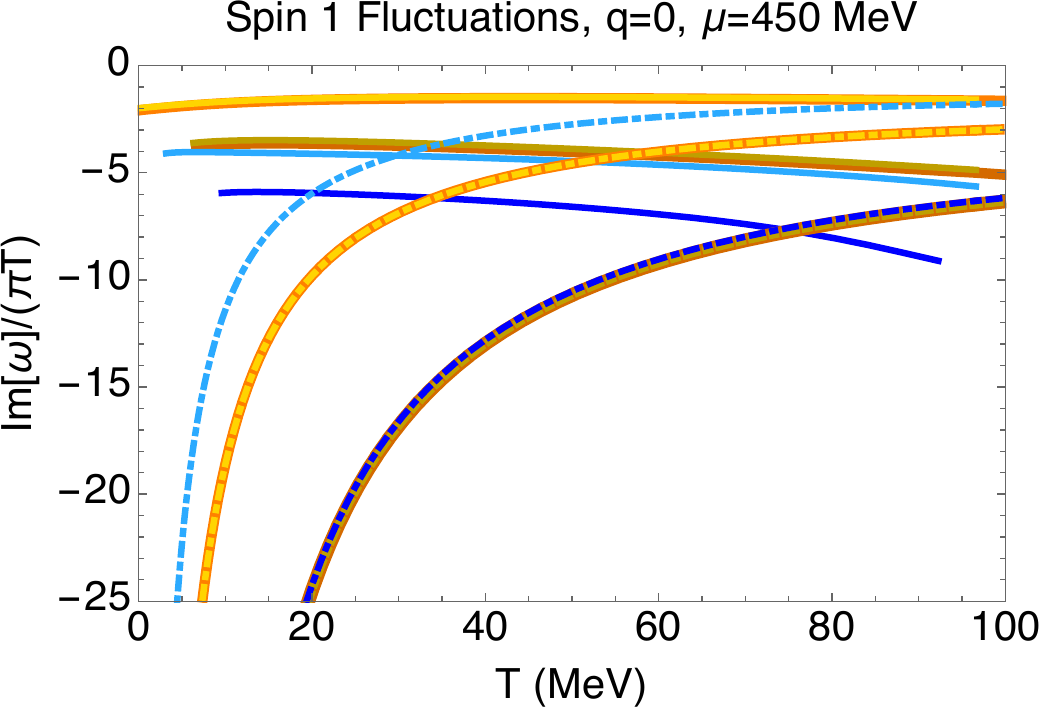}
 \includegraphics[width=7.3cm]{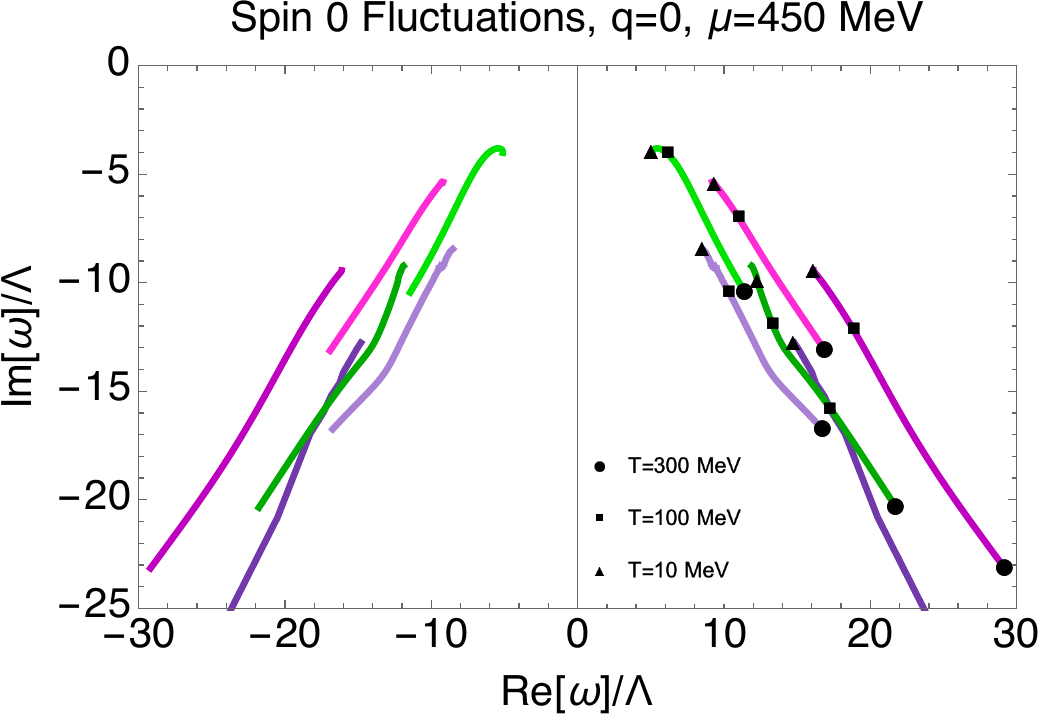}%
 \includegraphics[width=7.3cm]{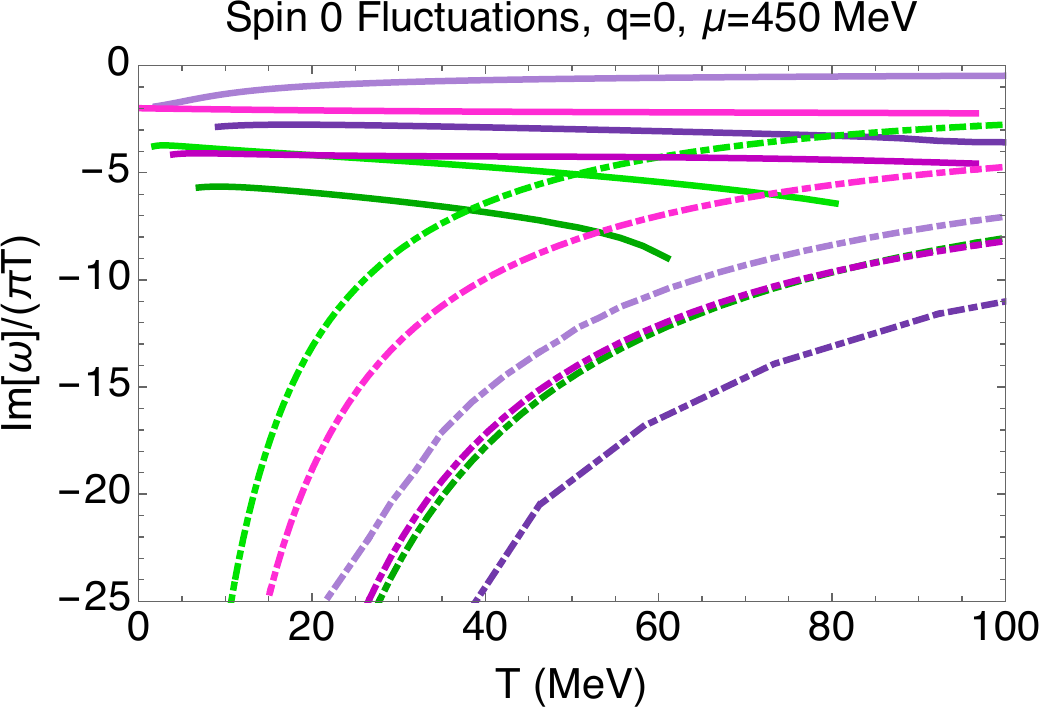}     
\caption{\small{Temperature dependence of QNMs at fixed chemical potential ($\mu=450$~MeV). The left column illustrates the temperature-dependent trajectories of the lowest two complex modes for each type of fluctuation. In the right column, the solid curves represent the temperature dependence of purely imaginary modes, while the dotdashed curves depict the imaginary part of the complex modes. The colors refer to fluctuations in different sectors as indicated in the legend, where NA stands for non-Abelian. Lighter (darker) shade shown the lowest (second to lowest) mode.}}
 \label{fig:allmodesq=0mu450}
\end{figure}

\begin{figure}[!tb]
\centering
\includegraphics[width=12cm]{plotlegend.pdf}
 \includegraphics[width=7.3cm]{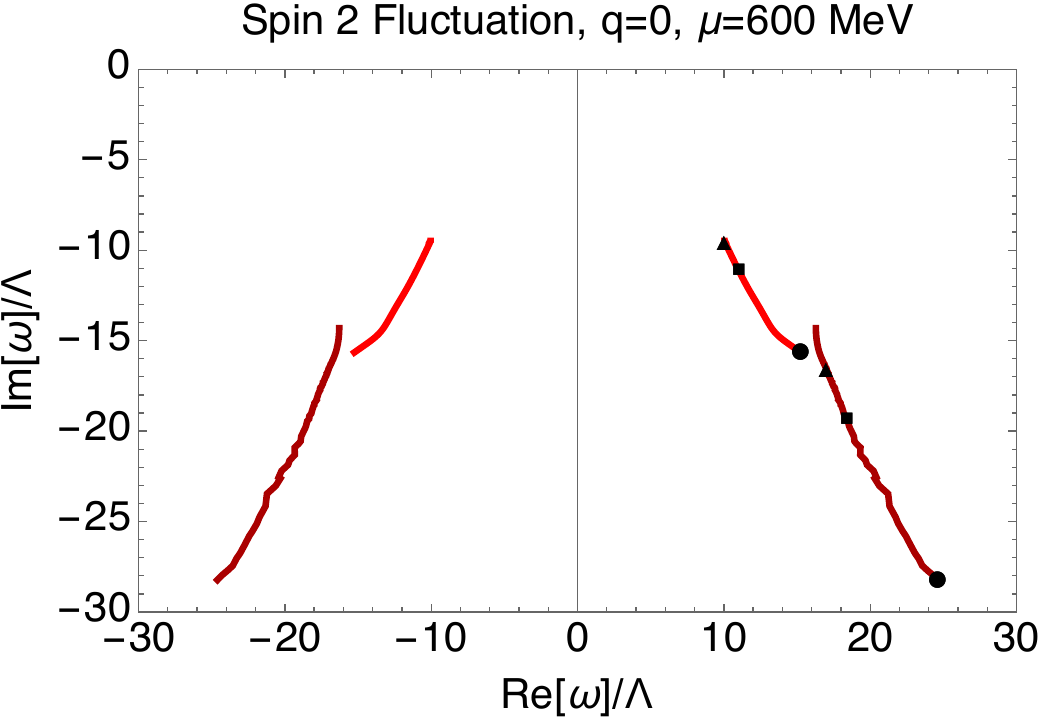}%
 \includegraphics[width=7.3cm]{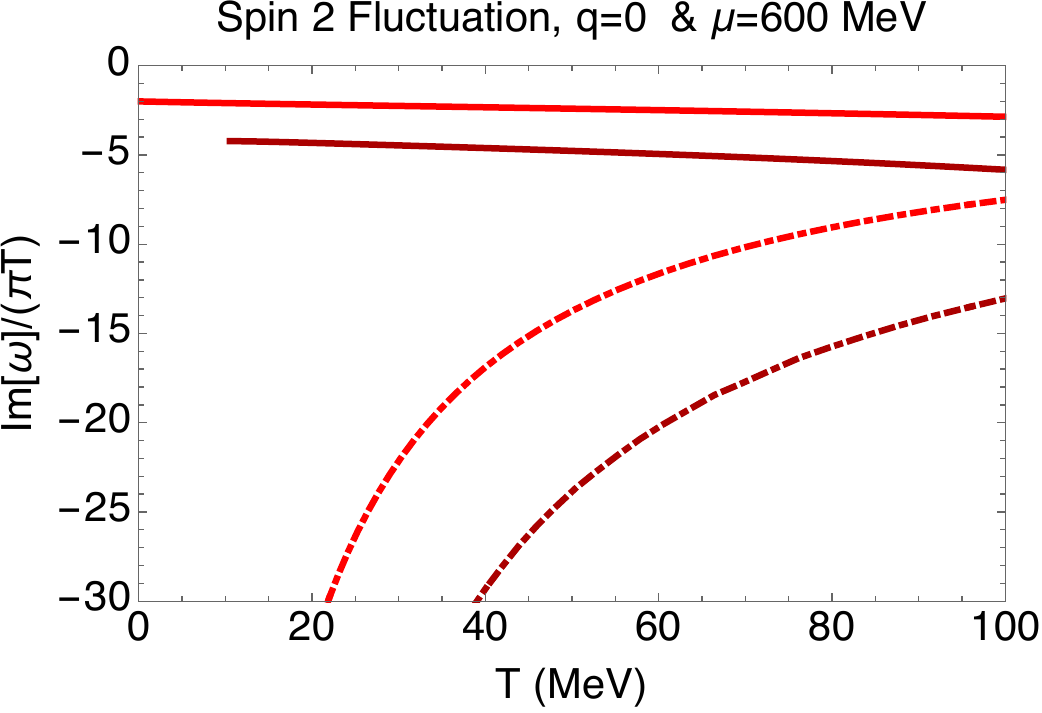}    \includegraphics[width=7.3cm]{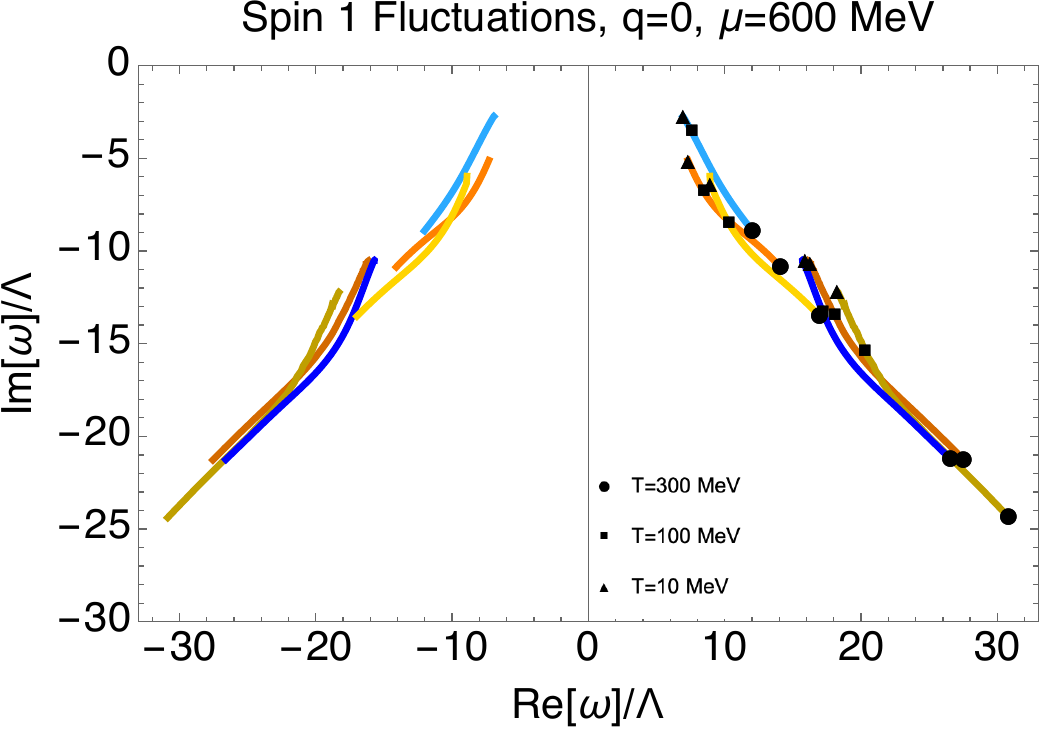}%
 \includegraphics[width=7.3cm]{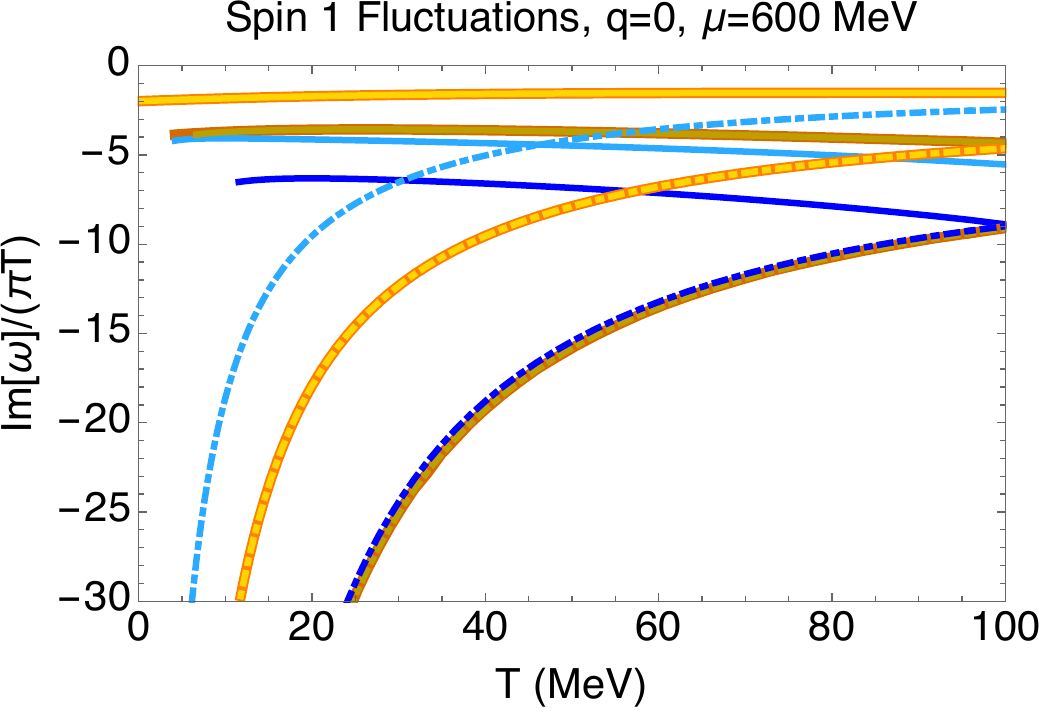}
 \includegraphics[width=7.3cm]{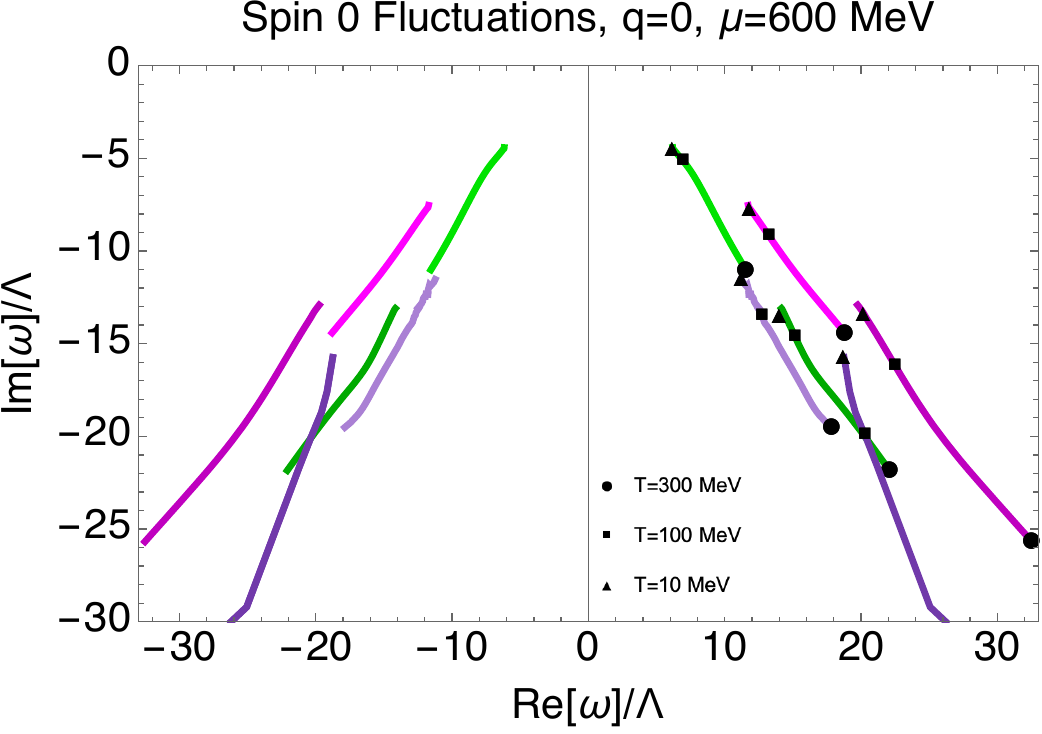}%
 \includegraphics[width=7.3cm]{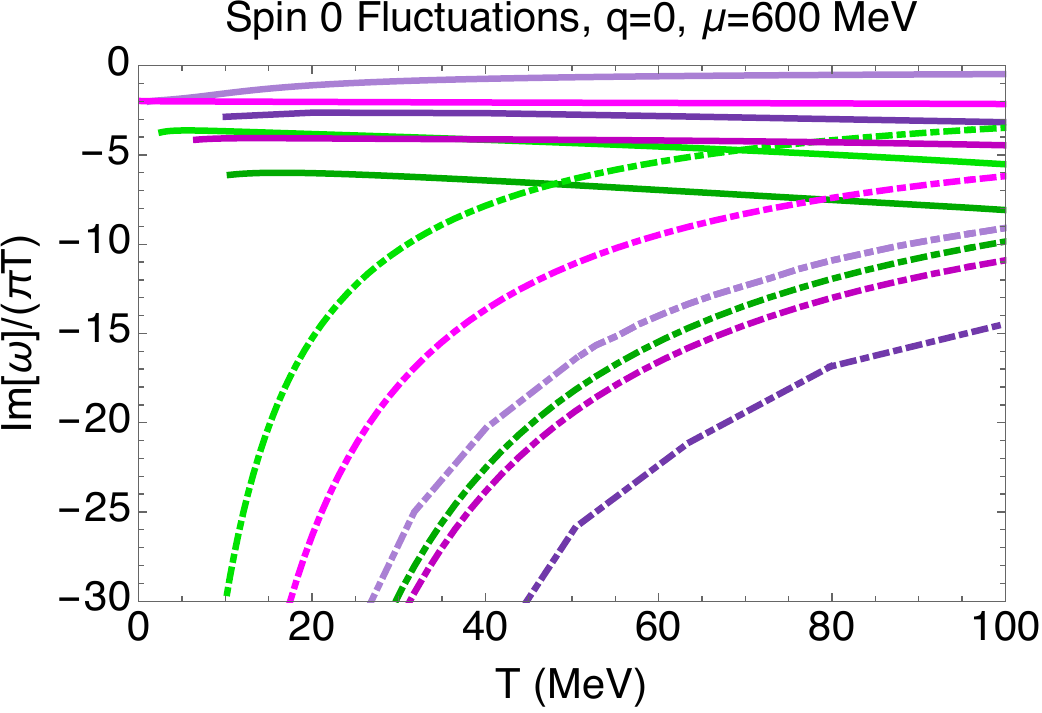}     
\caption{\small{Temperature dependence of QNMs at fixed chemical potential ($\mu=600$~MeV). Notation as in Fig.~\protect\ref{fig:allmodesq=0mu450}.}}
 \label{fig:allmodesq=0mu600}
\end{figure}

\subsection{Numerical analysis: temperature dependence 
 of QNMs}

In this subsection, we present the trajectories of QNMs obtained by numerically solving the fluctuation equations for each sector at zero momentum while varying the temperature.
The numerical solutions are found by using a variant of pseudospectral methods, which is described in Appendix~\ref{app:numerics}. 
The chemical potential is held fixed, and we investigate the temperature dependence for two different values: $\mu=450$ MeV and $\mu=600$ MeV. These values were chosen to lie above the critical chemical potential of the confinement-deconfinement transition, which is at $\mu_c \approx 406$~MeV for potentials 7a, but quite close to the transition in order to avoid the weakly coupled region at asymptotically high $\mu$ where holography is not expected to work.

As previously mentioned, at vanishing momentum, all modes decouple from each other and are classified according to spin rather than helicity.
The results for $\mu=450$ MeV and $\mu=600$ MeV are shown  
in Figure~\ref{fig:allmodesq=0mu450} and Figure~\ref{fig:allmodesq=0mu600}, respectively. For these plots, we used potentials 7a discussed in Sec.~\ref{sec:pot_choice}. 
In the left panels, we show the trajectories of the lowest two complex modes for each sector in the complex frequency plane. The curves are parameterized with temperature, and the locations of the QNMs at specific temperature values ($T=300$ MeV, $T=100$ MeV, and $T=10$ MeV) are marked on each curve. In this set of plots, both the imaginary and real parts of the QNM frequency are normalized with the energy scale $\Lambda 
\sim\Lambda_{QCD}$ 
to emphasize the temperature dependence of the trajectories. 
All the QNMs have 
similar 
temperature dependence. 
The detailed shape of the curves does however slightly depend on the mode, on the temperature, and on the chemical potential. The modes at higher $\mu$ (Fig.~\ref{fig:allmodesq=0mu600}) are typically further away from the real axis, in units of $\Lambda$, as the modes at lower $\mu$.

In the right panels of Figure~\ref{fig:allmodesq=0mu450} and Figure~\ref{fig:allmodesq=0mu600}, we show the temperature dependence of the purely imaginary modes with solid curves. Additionally, the imaginary parts of the complex modes are shown in the plots via dotdashed curves. In contrast to the right-hand side plots, the frequencies are normalized with $(\pi T)$ in these plots so that the temperature dependence of the imaginary modes is roughly divided out. The curves of some of the imaginary modes end abruptly before reaching $T=0$ because we cut out parts of the curves with significant numerical issues.

As one can see 
from the plots, pure imaginary modes and imaginary parts of the complex modes  
show the behavior expected for charge black holes: the frequencies of the imaginary modes are roughly proportional to the temperature, whereas the complex modes are roughly independent of the temperature \cite{Horowitz:1999jd}.
Due to this behavior, the complex modes give highly suppressed corrections 
to the dynamics at low temperatures. 
The lowest mode, which controls the equilibration process, remains purely imaginary within the temperature range considered. These modes are gravity, non-Abelian gauge, and axion modes for the spin two, spin one, and spin zero sectors, respectively. Overall, the mode with the longest lifetime is the leading, imaginary axion mode.

\begin{figure}[!tb]
\centering
\includegraphics[height=0.05\textwidth]{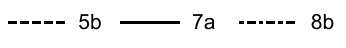}
\includegraphics[height=0.05\textwidth]{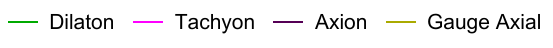}
\includegraphics[height=0.32\textwidth]{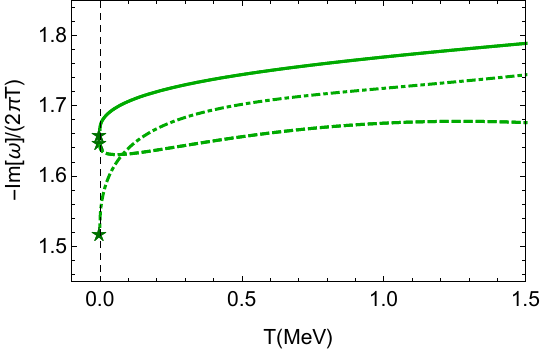}
\includegraphics[height=0.322\textwidth]{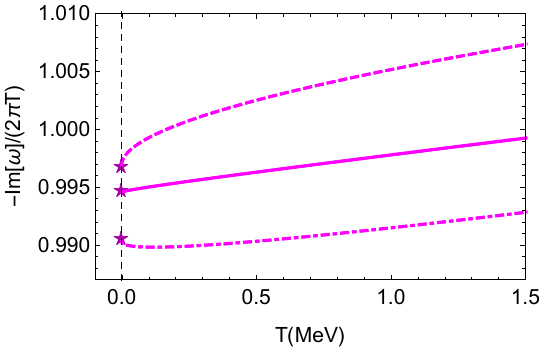}
\includegraphics[height=0.322\textwidth]{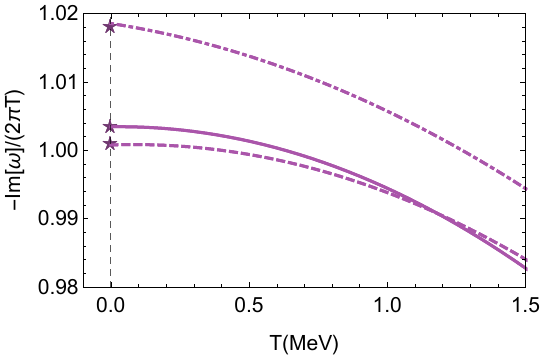}
\includegraphics[height=0.32\textwidth]{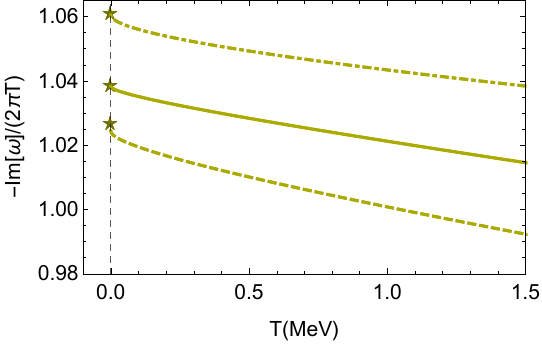}
   \caption{\small The numerical results at $\mu=450$~MeV for the nonuniversal dimensions at the AdS$_2$ fixed point. Stars mark the values given by the analytical expressions in the Table~\ref{tab:dimensions}. The results demonstrates the consistency of analytic and numerical calculation in the low-temperature and high-density regime.}
    \label{nonuni}
\end{figure}

Another 
observation is that the behavior of the imaginary 
modes at low temperatures is consistent with the formula~\eqref{Ads2qnm} and the AdS$_2$ dimensions computed in Sec.~\ref{ads2}.
From the plots in Figs.~\ref{fig:allmodesq=0mu450} and~\ref{fig:allmodesq=0mu600} it is however difficult to read off the low temperature limit precisely due to limited resolution.
To verify the agreement precisely, 
we therefore 
zoom into the behavior of QNMs for vanishing momentum and at 
low temperatures (down to $T=0.01$ MeV), very close to the AdS$_2$-critical line, focusing on the nonuniversal modes, i.e., the modes with noninteger AdS$_2$ dimensions. For this computation, the chemical potential is fixed to $\mu=450$~MeV. 

The results are shown in Fig.~\ref{nonuni}, where we use the same color code as in Figs.~\ref{fig:allmodesq=0mu450} and~\ref{fig:allmodesq=0mu600}: 
dilaton, tachyon, axion and Abelian axial gauge field modes are shown by the green, magenta, purple and yellow curves, respectively. Additionally, different styles of curves are used to represent different potentials: dashed, solid, and dotdashed curves correspond to 5b, 7a, and 8b potentials. The stars in the plots mark the values given by the analytical expressions presented in Table~\ref{tab:dimensions}.

In the top-left plot, we observe that the dilaton mode converges to the value predicted by the equation~\eqref{dildeltaflow} rather than the equation~\eqref{dildeltaads}. The remaining plots in Figure~\ref{nonuni} 
confirm that other nonuniversal modes also converge to the analytically computed values presented in Table~\ref{tab:dimensions}. The curves were obtained by fitting numerical data with linear combinations of (half) integer powers of the temperature. 
This Ansatz gives essentially a perfect fit to all the data.
The mean-field like $\sim\sqrt{T} + \mathrm{const}$ behavior  of the data at low temperature is clear from all other sectors except for the axion.

\section{QNMs at finite momentum}\label{qnmsfiniteq}

In this section, we explore numerically the momentum dependence of QNMs. We begin by analyzing the helicity one channels, where the CS term contributes and therefore modulated 
instabilities can arise~\cite{Domokos:2007kt,Nakamura:2009tf}. In Sec.~\ref{striped} and Sec.~\ref{subsechel1}, we conduct a comprehensive analysis of these instabilities. Subsequently, in Sec.~\ref{hel0s}, we examine the helicity-zero channel 
in the low-temperature and high-density regime. 

\begin{figure}[!tb]
\centering
\includegraphics[width=0.49\textwidth]{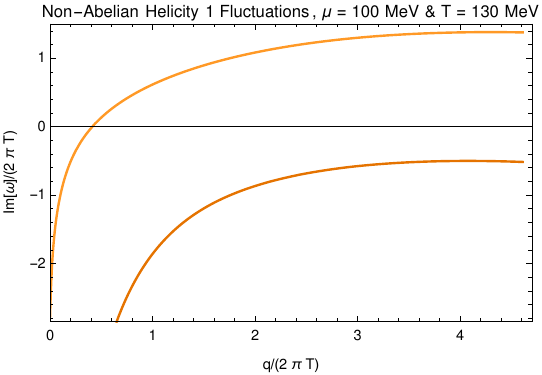}%
\hspace{2mm}\includegraphics[width=0.49\textwidth]{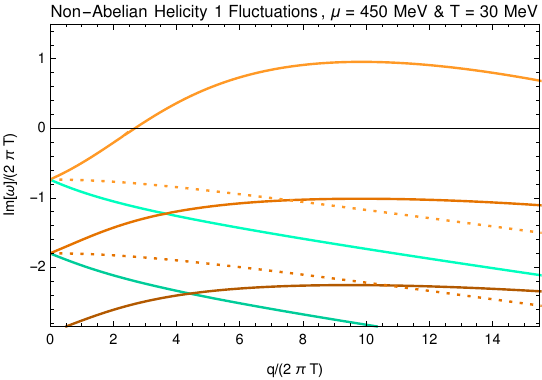}    

\vspace{2mm}

\includegraphics[width=0.49\textwidth]{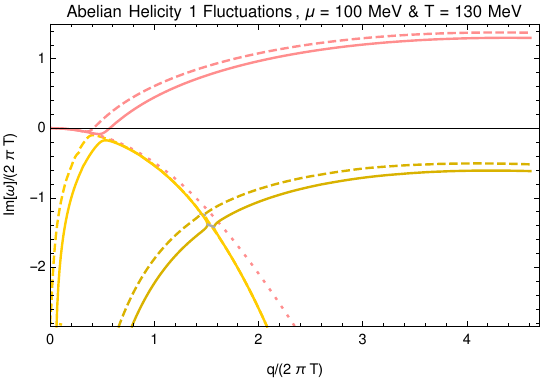}%
\hspace{2mm}\includegraphics[width=0.49\textwidth]{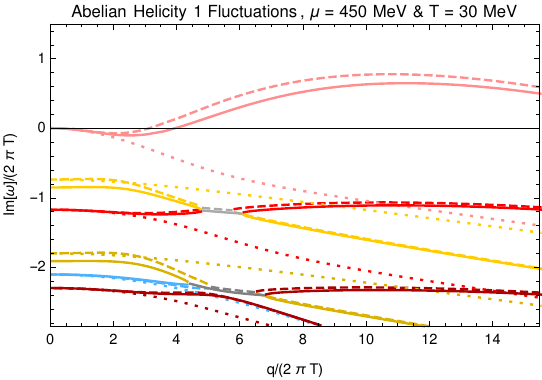}
    \caption{\small The momentum dependence of the QMNs in the non-Abelian (top row) and the Abelian (bottom row) subsectors of helicity one channel at $\mu=100$~MeV, $T=130$~MeV (left column) and $\mu=130$~MeV, $T=30$~MeV (right column). Different styles of the curves represents different situations: solid, dashed, dotted curves denote the full V-QCD model, the model without the $S_a$ term in~\protect\eqref{Sadef}, and the model without both the CS term in~\protect\eqref{eq:SCSdef} and the $S_a$ term, respectively. The colors of the imaginary modes on the top row refer to the nature of the mode at zero momentum as indicated in Figs.~\protect\ref{fig:allmodesq=0mu450} and~\protect\ref{fig:allmodesq=0mu600}. The gray curves are imaginary parts of complex modes. On the bottom mode, the orange (cyan) curves show the decoupled left (right) handed modes or vice versa, depending on the sign of helicity. The 
shade 
of the specific color refers to the location on the imaginary frequency axis, such that the mode with longest lifetime if brightest.
}
    \label{fig:hel1T30}
\end{figure}

\subsection{Modulated instability}\label{striped}

We  start by studying 
the spatially modulated instabilities  
in the Abelian and non-Abelian subsectors of the helicity one channel. 
Both subsectors are  
affected by the 
CS term.
We first analyze these instabilities in two different points 
of the phase diagram: 
the plots in the left and right columns of Fig.~\ref{fig:hel1T30} correspond to $\mu=100$ MeV, $T=130$ MeV, and $\mu=450$ MeV, $T=30$ MeV, respectively. The top row illustrates the non-Abelian subsector, while the bottom row displays the Abelian subsector.

In the non-Abelian subsector  (the top row in Fig.~\ref{fig:hel1T30}) the results have slightly simpler structure than in the Abelian sector (bottom row). Recall that the left and right handed modes are decoupled, and we use cyan and orange for these modes. Which of the curves is the left or right handed mode depends on the signature of the helicity. 

The most striking feature in these plots is the strong modulated instability: the lowest mode crosses to positive $\mathrm{Im}\,\omega$ at finite momentum. As the non-Abelian subsector only contains gauge field fluctuations, the dynamics is analogous to that of the Einstein-Maxwell-dilaton model where this effect was originally observed~\cite{Nakamura:2009tf,Ooguri:2010kt}. The details however depend on the location $(\mu,T)$ on the phase diagram: this is why we chose two different points for the plots in the left and right columns in Fig.~\ref{fig:hel1T30}. In the bottom left plot, the cyan modes are at such high negative $\mathrm{Im}\,\omega$ that they are not visible. We also note that if the CS term is turned off by hand (dotted curves in the bottom right plot) the instability disappears, as expected.

The main difference between the Abelian and non-Abelian sectors is the mixing of the gauge fields with the fluctuations of the metric in the Abelian sector. The gravity modes in this sector include the hydrodynamical shear mode, for which $\omega \to 0$ as $q \to 0$ (see the top row plots in Fig.~\ref{fig:hel1T30}). Consequently this mode is the most important mode at low momentum.

In the Abelian sector (the bottom row in Fig.~\ref{fig:hel1T30}), a qualitatively similar instability is observed as in the non-Abelian sector. In these plots, the colors indicate the classification of the modes in the limit of zero momentum, $q \to 0$, in the notation of Figs.~\ref{fig:allmodesq=0mu450} and~\ref{fig:allmodesq=0mu600}. We stress that as $q$ grows and the modes start to mix, the colors do not necessarily have anything to do with the nature of the mode, and have been added just to guide the eye. The similarity of the dispersion relations of the unstable modes between the Abelian and non-Abelian sectors however suggests that the unstable modes are dominantly left or right handed gauge fields also in the Abelian sector.

The dashed curves on the bottom row show the results in the absence of the $S_a$ term of~\eqref{Sadef}, which implements the physics of the axial U(1) anomaly.
The presence or absence of $S_a$ leads to a small effect, and there is no qualitative differences between the curves with or without $S_a$.
Turning off the term slightly enhances 
the instability.
We note that again that also turning off the CS term~\eqref{eq:SCSdef} (dotted curves) removes the instability, as expected.

\begin{figure}[!tb]
\centering
\includegraphics[height=0.35\textwidth]{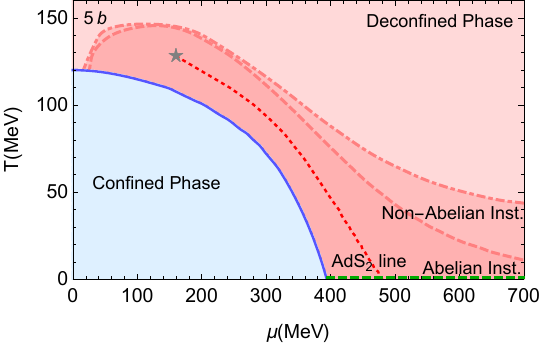}
\includegraphics[height=0.35\textwidth]{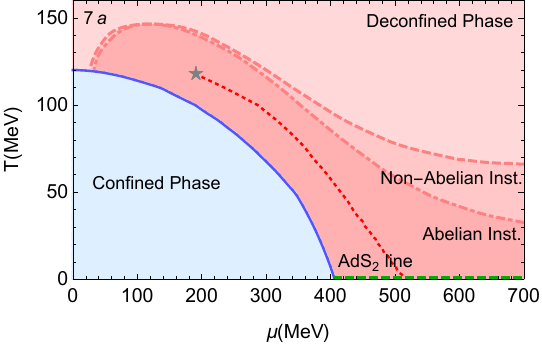}
\includegraphics[height=0.35\textwidth]{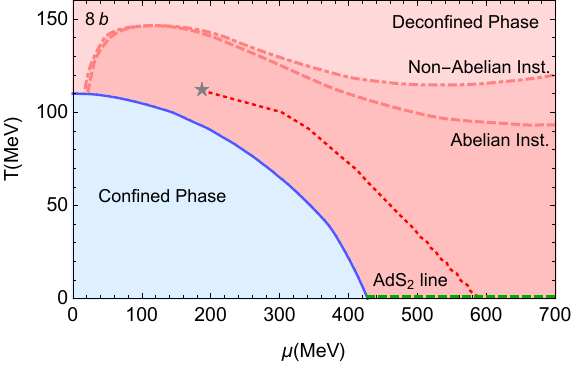}
   \caption{\small The phase diagrams of V-QCD, illustrating the instability analysis for different potential sets: 5b, 7a, and 8b, are presented in the top, middle, and bottom plots, respectively. The shaded region below the dashed (dot-dashed) curves indicates the region where the Abelian (non-Abelian) sector accommodates spatially modulated instabilities. The red dotted curves and stars denote the nuclear to quark matter transition and its critical endpoint estimated using a different approach in~\cite{Demircik:2021zll}. }
    \label{insfig}
\end{figure}

In this rest of this subsection, we 
investigate the extent of the instabilities 
using three different 
V-QCD potentials, namely 5b, 7a, and 8b. 
The results are presented in Fig.~\ref{insfig}, with the top, middle, and bottom plots corresponding to potentials 5b, 7a, and 8b, respectively. In each plot, the solid blue curves shows the boundary between the deconfined and confined phases.
The shaded region in the deconfined phase below the dashed (dot-dashed) curves indicates the values of temperature and quark chemical potential at which the Abelian (non-Abelian) sector exhibits the instability. 

The shape of the region where the instability appears is somewhat surprising: an instability at low temperature near the AdS$_2$ line (the dashed green curve along the horizontal axis) is expected, but the extent of the instability to low chemical potentials and relatively high temperatures is not. Recall that in the Reissner-Nordstr\"om background the onset of the instability only depends on the ratio $\mu/T$~\cite{Nakamura:2009tf}. 
Many features are common to all three potentials. Firstly, the instability curves peak at $T \sim 150$~MeV 
in the low-density regime 
and decrease towards higher chemical potential values. 
Secondly, the instability region of the non-Abelian sector is consistently larger than that of the Abelian sector.
The precise extent of the instability, in particular at high chemical potentials, however depends on the potentials: 
the region with instability is at its largest (smallest) for potentials 8b (5b). 

As it turns out, the surprising shape of the region with instability is closely related to our choice for the function $w(\phi)$, i.e., the coupling of the gauge fields in the DBI action~\ref{generalact}(see~\eqref{Senaction}). This can be read off from the fluctuation equation for the non-Abelian fluctuations~\eqref{eq:LRinstabilityeq}: Here the last term, which arises from the CS term, provides an effective negative mass squared contribution (for negative helicity) and drives the instability. Notably, this term is proportional to $w(\phi)^{-4}$, and consequently sensitive to the functional form of $w(\phi)$. We use here the function $w(\phi)$ fitted to the lattice data in~\cite{Jokela:2018ers}. More precisely, it is the baryon number susceptibility which determines the functional form of $w(\phi)$. Lattice studies indicate (see, e.g.,~\cite{Borsanyi:2011sw}) that the susceptibility decreases significantly with decreasing temperature, which implies that the fitted $w(\phi)$ decreases fast with $\phi$. Thus, for such solutions where large values of $\phi$ are reached, the instability is strongly enhanced. Since the value of the dilaton typically grows fast towards the IR, this means small black holes. Actually, the black hole solutions have been explored as a function of the horizon value of the dilaton in~\cite{Alho:2012mh,Alho:2013hsa,Hoyos:2021njg}, and it is the black holes with small chemical potentials and temperatures close to the critical temperature, for which highest horizon values of the dilaton are reached. 

Apart from $w(\phi)$, the term driving the instability in~\eqref{eq:LRinstabilityeq} is proportional to the charge $\hat n$, so it vanishes at zero chemical potential where the charge is also zero. The shape of the region with instability in Fig.~\ref{insfig} can therefore be understood to arise as the combination of these two effects: enhancement of the instability with growing charge and with growing dilaton values at the horizon.

We also compare our results 
to the transition line between the nuclear matter (NM) and quark matter (QM) phases, as computed in \cite{Demircik:2021zll}, which is shown as the red dotted curve. 
It is important to 
note that this transition line is calculated for the extended version of the model, which includes the NM phase and the pressure of electrons, and is included here solely for the purpose of comparison. 
The gray stars represent the critical end points 
from the same study. 
Interestingly, our results suggest that the critical point is ``cloaked'' by the instability independently of the details of the potentials.

The  
value of momentum (not shown in the plots) $q_c$ where the instability is at its strongest, i.e., $\mathrm{Im}\,\omega$ reaches its maximum, is of the order of $\Lambda$ for all values of $\mu$ and $T$ where the instability is found.
The precise value depends on the 
location in the phase diagram but the dependence on $T$ and $\mu$ is weak and the value of the momentum is always comparable to $\Lambda$. For example, in the low-density region (left panels of Fig.~\ref{fig:hel1T30}) we find that $q_c \approx 3.5$~GeV which corresponds to the wavelength $\sim 0.35$~fm for the modulation. In this region, the maximal value of $\mathrm{Im}\,\omega$ is $\sim 1.1$~GeV, corresponding to a time scale $1/(\mathrm{Im}\,\omega)$ of $\sim 0.2$ fm/$c$.

\begin{figure}[!tb]
\includegraphics[width=0.49\textwidth]{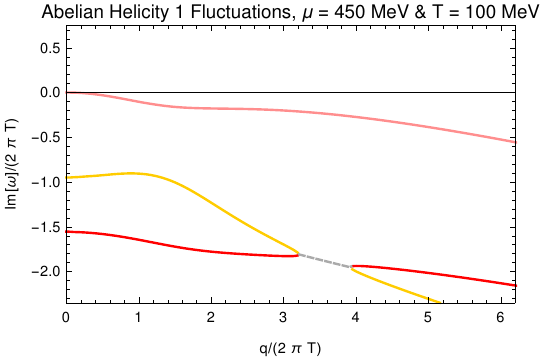}\hspace{3mm}%
\includegraphics[width=0.49\textwidth]{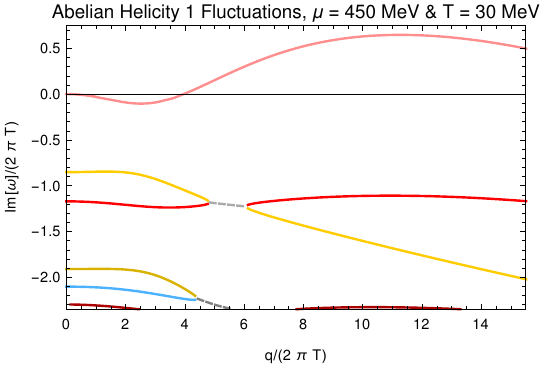}
\includegraphics[width=0.49\textwidth]{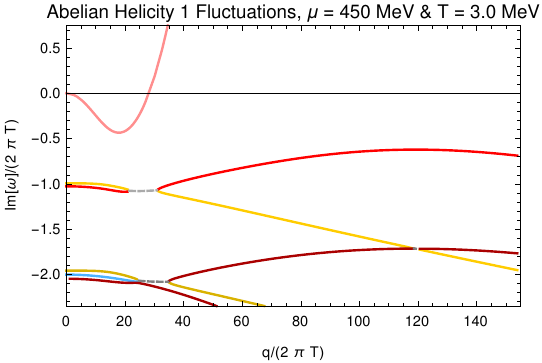}\hspace{3mm}%
\includegraphics[width=0.49\textwidth]{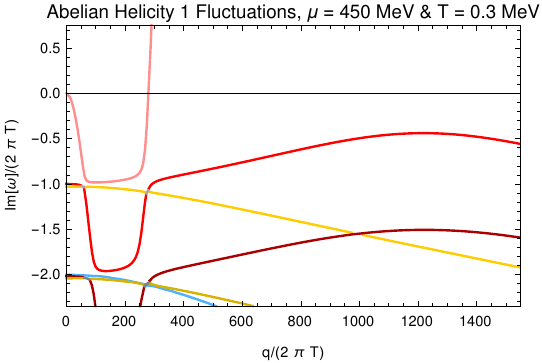}
    \caption{\small Momentum dependence of the modes in the Abelian subsector of helicity one channel at $\mu=450$~MeV and $T=100$~MeV, $T=30$~MeV, $T=3$~MeV, $T=0.3$~MeV. The color code shows the nature of the modes at $q=0$ as in the bottom row of Fig.~\protect\ref{fig:hel1T30}.
    }
    \label{fig:hel1Tdep}
\end{figure}

\subsection{Low temperature behavior in the helicity one sector}\label{subsechel1}

In this subsection, we focus on the Abelian subsector of the helicity one channel, and the behavior as we lower the temperature, approaching the AdS$_2$ line. In particular, we study how the instability affects the behavior of the hydrodynamic shear diffusion mode near the AdS$_2$ region~\cite{Arean:2020eus}. 
To this end, we analyze the momentum dependence at a fixed chemical potential slice of $\mu=450$ MeV while varying the temperature through four different values: $T=100$ MeV, $T=30$ MeV, $T=3$ MeV, and $T=0.3$ MeV. We use here potentials 7a.

The results are 
shown  
in Fig.~\ref{fig:hel1Tdep}, employing the same color code as section~\ref{qnm0q} as a function of the nature of the modes at zero momentum: red, blue and yellow represent the gravity, Abelian vector and Abelian axial modes respectively, with shade varying according to their location on the imaginary axis. We remind that this classification works only in the limit $q \to 0$. As $q$ grows past the value where clear mixing effect set in, the colors are only useful to guide the eye.

First we note that the behavior of the instability in Fig.~\ref{fig:hel1Tdep} is in agreement with Fig.~\ref{insfig}: At $T=100$~MeV there is no instability, but as the temperature is lowered, the effect of the CS term becomes more prominent, and the instability sets in. The unstable mode is not shown in the bottom row of Fig.~\ref{fig:hel1Tdep} because it is at considerably higher $\mathrm{Im}\,\omega$ than the scale of the imaginary AdS$_2$ modes. The maximal frequency of the unstable mode is roughly independent of the temperature, $\mathrm{Im}\,\omega \sim T^0$, as $T \to 0$, while the scale of the frequency of the AdS modes is set by the temperature, in agreement with Eq.~\eqref{Ads2qnm}. Similarly, the characteristic momentum of the instability (the value of $q$ at which $\mathrm{Im}\,\omega$ reaches its maximum) is roughly independent of the temperature. Notice that the scale of the plotted $q/(2\pi T)$ varies between the plots in Fig.~\ref{fig:hel1Tdep}. 

At sufficiently low temperatures, as displayed in the plots in the bottom row 
of Figure~\ref{fig:hel1Tdep}, the system enters the AdS$_2$ region and starts exhibiting ``universal'' behavior at low momentum~\cite{Arean:2020eus}. First, in agreement with the low temperature behavior in Figs.~\ref{fig:allmodesq=0mu450},~\ref{fig:allmodesq=0mu600}, and~\ref{nonuni}, the ratios $\mathrm{Im}\,\omega/(2\pi T)$ become integers up to small corrections. That is, the behavior follows the relation of Eq.~\eqref{Ads2qnm}. This appears to hold true even for the nonuniversal axial mode (yellow curves as $q \to 0$) which has noninteger $\Delta_*$, because the numerical value $\Delta_* \approx 1.0382$ is so close to one, the the difference cannot separated from temperature-dependent corrections to this formula. Second, the shear modes follows the standard dispersion relation $\omega \approx -iq^2$ in the relevant region as $T \to 0$, where the diffusion constant $D \sim 1/T$.

In the absence of the instability, the collision of the shear mode with the lowest AdS$_2$ modes (which occurs at complex momentum values) would set the radius of convergence of the hydrodynamic dispersion relation~\cite{Grozdanov:2019kge,Grozdanov:2019uhi}, and determine the characteristic equilibration time of the system. In the limit of $T \to 0$ the result is universal~\cite{Arean:2020eus}: it is determined in terms of the diffusion constant and the dimension of the leading AdS$_2$ mode. This happens because, as one can see from Fig.~\ref{fig:hel1Tdep}, the scale of the momentum dependence of the AdS$_2$ modes is $\sim T^0$. Therefore the AdS$_2$ modes have not yet started to move at the momentum $q \sim T$ where the collision takes place, as we take $T \to 0$. The higher order corrections in $q$ to the dispersion relation of the shear mode are also suppressed for $q \sim T$. This leads to the result $\omega_\mathrm{coll} \approx 2\pi T$, $q_\mathrm{coll} \approx \sqrt{2\pi T}$ for the location of the collision. In particular, the collision happens at almost real momentum.
Moreover, the AdS$_2$ modes and the shear mode are essentially decoupled as $T \to 0$ as one can see from the bottom right plot in Fig.~\ref{fig:hel1Tdep}: the shear mode passes the AdS$_2$ modes with suppressed interaction. This agrees with the observation in~\cite{Davison:2013bxa} that unlike suggested by the location of the collision, the behavior of hydrodynamics in the region is controlled by the chemical potential rather than temperature. 

We however stress that in the presence of a perturbative instability, such as in the V-QCD model, the discussion of the  features of hydrodynamics is academic. For the collision of the modes to have a clear physical meaning, we would need to 
turn off the CS term by hand.  This would indeed remove the instability, whereas the low momentum structure of the  bottom right plot in Fig.~\ref{fig:hel1Tdep} would still remain. Nevertheless, it is interesting to see that the low-temperature behavior of the QNMs in V-QCD is essentially the same as in the neutral translation-breaking AdS$_4$ and AdS$_4$-Reissner-Nordstr\"{o}m backgrounds studied in~\cite{Arean:2020eus}. That is, the current model is much more complex than in the earlier study, and even includes the instability driven by the CS term, but the picture at low momentum is unaffected by this.

\begin{figure}[!tb]
\center{\includegraphics[width=0.7\textwidth]{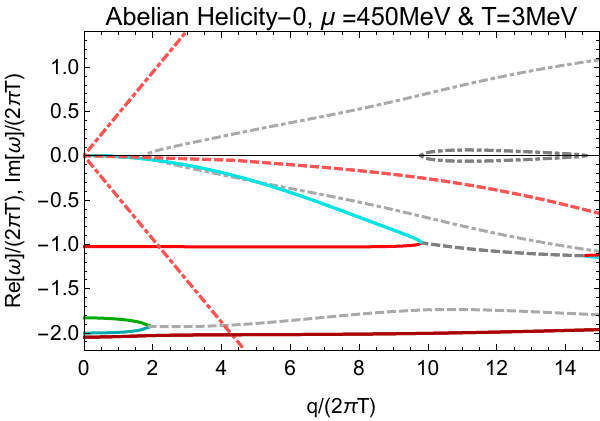}}
    \caption{\small The plot for momentum dependence of the Abelian subsector of helicity-zero channel at $\mu=450$~MeV, $T=3$~MeV. While  solid curves represent the purely imaginary modes, dashed (dotdashed) depicts imaginary (real) parts of the complex mode. The colors show the nature of the modes at $q=0$ following the notation in Figs.~\protect\ref{fig:allmodesq=0mu450} and~\protect\ref{fig:hel1T30}.
    }\label{hel0qdep}
\end{figure}

\subsection{Low temperature behavior in the helicity zero sector}\label{hel0s}

In this subsection, we conduct an analysis of the momentum dependence of the QNM spectrum in the helicity zero channel.
We focus on the $\{Z_0,\hat{E}_V,Z_\phi\}$-subsector of Fig.~\ref{finq_class} 
as it involves coupling between metric, Abelian vector gauge field and dilaton fluctuations, and therefore 
accommodates the sound and diffusion hydrodynamic modes. 

We specifically investigate the situation in the AdS$_2$ region,  
i.e., low temperature and high density. 
The momentum dependence of the 
spectrum in this 
region is shown 
in Figure~\ref{hel0qdep} 
at $T=3$ MeV and $\mu=450$ MeV. As above, 
the QNMs are denoted using a color code referring to their nature at low momentum: red, blue, and green represent modes identified as gravity, Abelian vector and dilaton modes at $q \to 0$, respectively. The shade  
of the curves corresponds to their location on the complex axis. The representation of modes is such that solid curves are used for purely imaginary modes, while the imaginary (real) parts of the complex modes are depicted by dashed (dot-dashed) curves.
 
In Fig.~\ref{hel0qdep}, the sound and charge diffusion hydrodynamic modes are represented by the  
light red dashed and  
light blue curves, respectively. For these modes 
$\omega\rightarrow 0$ as $q \rightarrow 0 $. 
As expected they are the most dominant modes in the low $q$-region and they have the expected momentum dependence: The real part of the sound mode is given  
by the red dot-dashed curves, having a linear dependence, while the 
imaginary part, as well as the purely imaginary diffusion modes, display a $q^2$ behavior as $q\rightarrow0$. Notably, the imaginary 
part of the 
frequency of the sound mode is smaller in magnitude than that of the diffusion mode. This indicates 
that model has low bulk viscosity. 

The modes identified as the metric and Abelian vector fluctuations as $q \to 0$ are represented by the darker red and darker blue solid curves, respectively. 
These modes are located along the imaginary axis 
near the values expected for universal AdS$_2$ modes when $q \to 0$, i.e., at integer $\mathrm{Im}\,\omega/(2\pi T)$. 
This observation confirms that the temperature (chemical potential) value that is considered is sufficiently low (high), such that the effects of the AdS$_2$ region can be observed. 
Moreover, the mode identified as the dilaton mode in the limit $q \to 0$ approaches 
a position consistent with the AdS$_2$ dimension calculated in Sec.~\ref{ads2}.

As it happens for the shear hydrodynamic mode in the Sec.~\ref{subsechel1} at low temperatures, 
the hydrodynamic diffusion mode interacts with the leading imaginary AdS$_2$ mode. The picture is again similar to the findings of
~\cite{Arean:2020eus} for neutral translation-breaking AdS$_4$ and AdS$_4$-Reissner-Nordstr\"{o}m geometries. 
In this case, the modes collide at real momentum, while for the shear mode and for the AdS$_4$ analysis of~\cite{Arean:2020eus} the collision takes place at a momentum value with small but nonzero 
imaginary part. 
We remark that the AdS$_2$ modes depicted in Fig.~\ref{hel0qdep} with dark blue and green solid curves also undergo a 
collision, resulting in the emergence of a 
pair of complex modes at a relatively low momentum. 

Notice that the value of the temperature ($3$~MeV) is significantly higher than the lowest value in Figure~\ref{fig:hel1Tdep}. Because of this, the decoupling of the hydrodynamic diffusion mode from the AdS$_2$ modes is less clear. We expect that at even lower temperatures (which we do not explore here due to numerical difficulties) the decoupling of these modes becomes more evident, leading to the momentum of the collision to be determined by the universal value as pointed out in~\cite{Arean:2020eus}.
Recall however that the low temperature region is perturbatively unstable due to the instability in the helicity-one sector, so this universal behavior seems to lack a physically meaningful interpretation in our case, unless one turns off the CS term giving rise to the instability by hand. Nevertheless it is again interesting that the crossing structure of the modes in the relatively complex V-QCD model is similar to what was found in the somewhat simple AdS$_4$ black holes in~\cite{Arean:2020eus}.

\acknowledgments

We thank for discussions D.~Arean, C.~Ecker, U.~G\"ursoy, T.~Ishii, N.~Jokela, M.~Kaminski, K.-S. Kim, E.~Kiritsis, S.~Nakamura, F.~Nitti, F.~Pe\~na-Ben\'\i{}tez, A.~Piispa, E.~Pr\'eau, M.~Roberts, D.~Schaich, and A.~Schmitt. 
T.D. 
acknowledges the support of the Narodowe Centrum Nauki (NCN) Sonata Bis Grant No.
2019/34/E/ST3/00405. J.~C.~R. and M.~J. have been supported by an appointment to the JRG Program at the APCTP through the Science and Technology Promotion Fund and Lottery Fund of the Korean Government. J.~C.~R. and M.~J. have also been supported by the Korean Local Governments -- Gyeong\-sang\-buk-do Province and Pohang City -- and by the National Research Foundation of Korea (NRF) funded by the Korean government (MSIT) (grant number 2021R1A2C1010834). J.~C.~R.  was additionally supported by
a DGAPA-UNAM postdoctoral fellowship.

\appendix

\section{The choice of the function \texorpdfstring{$Z(\phi)$}{TEXT}} \label{app:Z}

As discussed in the main text, we choose an Ansatz for the function $Z(\phi)$, appearing in the axion term~\eqref{Sadef}, as follows~\cite{Gursoy:2012bt}:
\be
 Z(\phi) = Z_0 \left[1 + c_a\, \frac{e^\phi}{8\pi^2} + d_a \left(\frac{e^\phi}{8\pi^2}\right)^4 \right] \ .
\ee
As it turns out, a nonzero value of $Z_0$ gives rise to an anomalous dimension for the axial current and axial chemical potential, given by~\cite{Arean:2016hcs,Weitz} 
\be
\gamma_A = -1 +\sqrt{1+\frac{4 N_f Z(-\infty)\ell^2}{N_c V_{f0}(-\infty)w(-\infty)^2}}  \ ,
\ee
where $Z(-\infty)=Z_0$ and $\ell$ is the UV AdS radius which satisfies
\be
 \ell^2 = \frac{12}{V_g(-\infty)- \frac{N_f}{N_c}V_{f0}(-\infty)} \ .
\ee 
As pointed out in~\cite{Weitz}, if $\gamma_A>1$, the axial current will be an irrelevant operator. In this case there is no solution of the geometry that would be asymptotically AdS$_5$ at finite axial chemical potential. To avoid this issue, we should require $\gamma_A<1$. This implies that 
\be
 Z_0 < \frac{3N_cV_{f0}(-\infty)w(-\infty)^2}{4N_f\ell^2} \ . 
\ee
For the potential sets 5b, 7a, and 8b that we use in this article, this bound evaluates to (see~\cite{Jokela:2018ers,Ishii:2019gta} for the precise definitions of the potentials)
\be \label{eq:Z0bounds}
 Z_0 < 2.116\quad (5b)\ , \qquad  Z_0 < 0.906\quad (7a)\ , \qquad Z_0 < 1.893\quad (8b) \ .
\ee

The function $Z(\phi)$ also determines the value of the (Yang-Mills) topological susceptibility, which is given as~\cite{Gursoy:2007er,Gursoy:2009jd,Gursoy:2012bt}
\be
 \chi_\mathrm{YM}^{-1} = M_p^{-3} \int_0^{\infty}\frac{dr}{e^{3A_\mathrm{YM}(r)}Z(\phi_\mathrm{YM})} \ ,
\ee
where $A_\mathrm{YM}$ and $\phi_\mathrm{YM}$ are the background functions for the pure glue theory, i.e., the model in the absence of flavors  ($N_f=0$). The value of the topological susceptibility and its dependence on $N_c$ has been studied extensively on the lattice~\cite{Lucini:2001ej,Bonati:2016tvi,Bonanno:2020hht,Athenodorou:2021qvs,Bennett:2022gdz}. Here we use the result $\chi_\mathrm{YM}/\sigma^2 = 0.01937(136)$ from~\cite{Bennett:2022gdz} extrapolated to $N_c=\infty$, where $\sigma$ is the string tension. Instead of computing the string tension in the holographic model, we study the dimensionless ratio $\chi_\mathrm{YM}/T_c^4$. The Yang-Mills deconfinement temperature $T_c$ can be computed in the model by determining the entropy and the temperature using black hole thermodynamics and integrating to obtain the free energy difference between the confined and deconfined phases~\cite{Gursoy:2008bu,Gursoy:2008za}. For the lattice value of the critical temperature, we use the $N_c=\infty$ result $T_c/\sqrt{\sigma} = 0.5970(38)$ from~\cite{Lucini:2005vg}, so that $\chi_\mathrm{YM}/T_c^4 \approx 0.152$.

Setting the coefficients $c_a$ and/or $d_a$ to small values (i.e., values less than one) and requiring agreement with the lattice results typically leads to values of $Z_0$ that violate the bounds~\eqref{eq:Z0bounds} roughly by an order of magnitude. However, as pointed out in~\cite{Weitz}, when these coefficients are increased, the value of $Z_0$ decreases rapidly. Here we choose the following values: $c_a=3$ and $d_a=20$, which were obtained by trial and error.\footnote{Alternatively, one could also impose an Ansatz for which $Z(-\infty) = 0$, in which case the UV anomalous dimension vanishes and the bound of~\protect\eqref{eq:Z0bounds} is absent.} For this choice, fitting to the lattice value of $\chi_\mathrm{YM}/T_c^4$ gives $Z_0 \approx 0.670$, which is below the bound for all potentials sets  5b, 7a, and 8b. Notice that the fitted value is the same for all the potential sets, because it only depends on the Yang-Mills sector, i.e., the function $V_g(\phi)$, which is the same in all these sets.

\section{Equations of motion}\label{app:eoms}

\subsection{General equations}

In this appendix we present the full equations of motion arising from the action \eqref{fullaction} in the case of vanishing tachyon, and
discuss some of their consequences.

The equations for the right and left handed gauge fields are given by
\begin{align}\label{gaugeeomR}
-\frac{M_p^3 N_c}{4}\partial_N \left(
   V_f(\phi(r))w(\phi(r))\sqrt{-\textrm{det}({\bf A}_{R})}\left( ({\bf A}_R^{-1})^{NM}-({\bf A}_R^{-1})^{MN} \right)\right)& \\
    -\frac{1}{24 \pi^2 M_p^3}\Big[ \partial_N\Big(A^R_{P}F^R_{QS}+\frac{i}{2} A^R_P A^R_Q A^R_S\Big)+\frac{1}{2}A^R_P A^R_Q A^R_S A^R_N \nonumber & \\
    -\frac{1}{4}F^R_{PQ} F^R_{SN}-\frac{3i}{4}A^R_P A^R_Q F^R_{SN}]\epsilon^{PNMQS} \nonumber & \\
    -M_p^3\sqrt{-\det g}\, Z(\phi)g^{MN}\left(N_c\partial_N \hat a - 
 \mathrm{Tr}\left(A^L_{N}-A^R_{N}\right)\right)=0\nonumber
\end{align}
and
\begin{align}\label{gaugeeomL}
-\frac{M_p^3 N_c}{4}\partial_N \left(
   V_f(\phi(r))w(\phi(r))\sqrt{-\textrm{det}({\bf A}_{L})}\left( ({\bf A}_L^{-1})^{NM}-({\bf A}_L^{-1})^{MN} \right)\right)& \\
    +\frac{1}{24 \pi^2 M_p^3}\Big[ \partial_N\Big(A^L_{P}F^L_{QS}+\frac{i}{2} A^L_P A^L_Q A^L_S\Big)+\frac{1}{2}A^L_P A^L_Q A^L_S A^R_N \nonumber & \\
    -\frac{1}{4}F^L_{PQ} F^L_{SN}-\frac{3i}{4}A^L_P A^L_Q F^L_{SN}]\epsilon^{PNMQS} \nonumber & \\
    +M_p^3\sqrt{-\det g}\, Z(\phi)g^{MN}\left(N_c\partial_N \hat a - 
 \mathrm{Tr}\left(A^L_{N}-A^R_{N}\right)\right)=0 \ ,\nonumber
\end{align}
respectively. The axion equation is given as
\be \label{axioneom}
 \partial_M \left[g^{MN} \sqrt{-\det g}\, Z(\phi) \left(N_c\partial_N \hat a -
 \mathrm{Tr}\left(A^L_{N}-A^R_{N}\right)\right) \right]=0
\ee
where as the dilaton equation reads
\begin{align}\label{dilatoneom}
\frac{8}{3 \sqrt{-\det (g)}}\partial_P\left( \sqrt{-\det (g)}g^{PQ}\partial_Q\phi \right)+\frac{\partial V_g(\phi)}{\partial \phi}& \\
-\frac{\partial \left( x V_{f0}(\phi)\sqrt{-\det({\bf A}_{R})}\right)}{\sqrt{-\det (g)}\partial \phi}-\frac{\partial \left( x V_{f0}(\phi)\sqrt{-\det({\bf A}_{L})}\right)}{\sqrt{-\det (g)}\partial \phi}=0 \ .\nonumber
\end{align}

The Einstein equations take the form
\begin{equation}
R_{M N}-\frac{1}{2} g_{M N} R=T_{M N}^g+T_{M N}^f+T_{M N}^a \label{EE}
\end{equation}
where the various contribution to the bulk energy-momentum tensor are given by
\begin{align}
&T_{M N}^g=\frac{-4}{3}\partial_M \phi \partial_N \phi-\frac{1}{2}g_{MN}V_g(\phi)+\frac{2}{3}g_{MN}\partial_P\phi \partial^P\phi\ & \\
&T_{M N}^f=\frac{1}{4}xV_{f0}(\phi)\frac{\sqrt{-\det({\bf A}_{R\,MN})}}{\sqrt{-\det(g_{MN})}}g_{MP}(({\bf A}^{-1}_R)^{PQ}+({\bf A}^{-1}_R)^{QP})g_{QN}+ \left(R\leftrightarrow L\right),\ & \\
& T_{M N}^a=Z(\phi) \Big[\frac{1}{2}\left(\partial_P \hat a - 
 \mathrm{Tr}\left(A^L_{P}-A^R_{P}\right)\right)^2g_{MN} \ \nonumber & \\
& -\left(\partial_M \hat a - 
 \mathrm{Tr}\left(A^L_{M}-A^R_{M}\right)\right)\left(\partial_N \hat a - 
 \mathrm{Tr}\left(A^L_{N}-A^R_{N}\right)\right)\Big] \ .
\end{align}

\subsection{Equations of motion for the background}

We then write down the equations of motion for the background. We use the metric in~\eqref{metric}, eliminate the Abelian vectorial field using~\eqref{nhatdef}, and set other gauge fields, the tachyon, and the axion to zero. The Einstein~\eqref{EE} and dilaton~\eqref{dilatoneom} equations imply that
\begin{align}\label{feq}
    f''+3 A'   f' &= \frac{1}{3} e^{2 A} \frac{\pa V_\mathrm{eff}(\phi,A,\hat n)}{\pa A} =\frac{x 
    \hat n^2e^{-4A}}{V_{f0}(\phi)w(\phi)^2R(\phi,A,\hat n)} \ ,& \\ 
    \label{Aeq}
A''-\left(A'\right)^2+\frac{4 }{9
   }\left(\phi '\right)^2&=0\ ,  &\\
 \frac{3 A' f'}{f}+12  
\left(A'\right)^2-\frac{4 }{3 }\left(\phi '\right)^2&=\frac{e^{2 A} V_\mathrm{eff}(\phi,A,\hat n)}{f} =\frac{e^{2A}}{f}\left[V_g(\phi) - x V_{f0}(\phi) R(\phi,A,\hat n)\right]\ , &\\
\phi '' + 3 A' \phi '+\frac{f' \phi '}{f} &= -\frac{3e^{2A}}{8f} \frac{\pa V_\mathrm{eff}(\phi,A,\hat n)}{\pa \phi}=\label{dileq}
&\\\nonumber 
&=-\frac{3e^{2A}}{8f}\left[V_g'(\phi)-\frac{xV_{f0}'(\phi)}{R(\phi,A,\hat n)}+\frac{x\hat n^2 e^{-6 A} w'(\phi)}{V_{f0}(\phi)w(\phi)^3R(\phi,A,\hat n) }\right] &
\end{align}
where the dependence on the potentials only appears through
\begin{align} \label{Veffdef}
 V_\mathrm{eff}(\phi,A,\hat n) &\equiv V_g(\phi) - x V_{f0}(\phi) R(\phi,A,\hat n) \ , &\\
  R(\phi,A,\hat n) &\equiv\sqrt{1+\frac{\hat n^2}{e^{6A}V_{f0}(\phi)^2w(\phi)^2}} \ .
\end{align}
This expression generalizes the effective potential of~\eqref{eq:Veffsimple} to finite density.
Notice that these equations are not independent: for example the dilaton equation~\eqref{dileq} can be derived by using the other equations. 

Apart from shifts in $r$, the equations are invariant under the following scale transformations:
\begin{align} \label{rtrans}
 r &\mapsto \Lambda_r^{-1}r \ ,& \quad A &\mapsto A +\log \Lambda_r \ ,& \quad \hat n &\mapsto \Lambda_r^3\hat n \ ; & \\
 f &\mapsto \Lambda_f^2 f \ ,& \qquad A &\mapsto A +\log \Lambda_f  \ ,& \quad \hat n &\mapsto \Lambda_f^3\hat n \ .&
 \label{ftrans}
\end{align}

\section{Fluctuation equations}\label{flucEq}
In this Appendix, we provide the linearized fluctuation equations for the GIVs. The equations build a system of second order coupled (decoupled) ordinary differential equations at finite 
(zero) momentum. The classification of fluctuations and coupling among GIVs are discussed 
in Sec.~\ref{flucs}. 
Since the  equations are complicated and their derivation is rather tedious, it is important to carry out consistency checks in order to make sure that the expressions are correct: 
 \begin{itemize}
  \item  After expressing the fluctuation equations in terms of GIVs, all non-gauge-invariant terms cancel out in the final equations. 
  \item All terms in the equations have consistent dimensions, and in more general, respect the symmetries~\eqref{rtrans}--\eqref{ftrans}. 
  \item In the small charge limit, the equations reduce to 
 the fluctuation equations of the Einstein-Maxwell-dilaton system, given as a part of the ``QNMspectral'' mathematica package~\cite{Jansen:2017oag}.
 \item  The equations reduce to previously known expressions~\cite{Arean:2013tja,Arean:2016hcs} in the limit of zero temperature and density.
 \item The equation decouple and respect the classification of modes presented in the main text in the limit of zero momentum. 
 \end{itemize}

 In the rest of Appendix, we present the system of the equations first at finite momentum in following three subsections, following the classification in Fig.~\ref{finq_class}. We then present the zero momentum limit of the equations (see Fig.~\ref{q0class} for the classification). Finally, we discuss the behavior of the dilaton fluctuation equation near the AdS$_2$ point in the last subsection. 

\subsection{Helicity two}

The helicity two sector only contains fluctuations of the metric which satisfy
\be
\left(Z_2^\pm\right)'' + \left(\frac{f'}{f}-A'\right)\left(Z_2^\pm\right)'+  \left(\frac{\omega ^2}{f^2}-\frac{q^2}{f}-\frac{2 A' f'}{f}-4 A'^2+\frac{8}{9} \phi '^2\right) Z_2^\pm = 0
\ee
or equivalently
\be
\left(e^{-2A}Z_2^\pm\right)'' + \left(3A'+\frac{f'}{f}\right)\left(e^{-2A}Z_2^\pm\right)'+  \left(\frac{\omega ^2}{f^2}-\frac{q^2}{f}\right) e^{-2A}Z_2^\pm = 0 \ .
\ee

\subsection{Helicity one}

The helicity one sector contains the non-Abelian gauge field equations
\begin{equation}
\begin{aligned}
&(\delta V^{\pm\, a})''+\left[\frac{d}{dr} \log e^{A}f w(\phi)^2V_{f0}(\phi)R\right](\delta V^{\pm\, a})'\\
&+\left[\frac{\omega^2}{f^2}-\frac{q^2}{f R^2}\right]\delta V^{\pm\, a}\pm \frac{e^{-2 A} \hat{n} q}{2 \pi^2 f M^3_p R^2 w(\phi)^4 V_{f0}(\phi)^2}\delta A^{\pm\, a}=0
\end{aligned}
\end{equation}
\begin{equation}
\begin{aligned}
&(\delta A^{\pm\, a})''+\left[\frac{d}{dr} \log e^{A}f w(\phi)^2V_{f0}(\phi)R\right](\delta A^{\pm\, a})'\\
&+\left[\frac{\omega^2}{f^2}-\frac{q^2}{f R^2}\right]\delta A^{\pm\, a} \pm \frac{e^{-2 A} \hat{n} q\delta }{2 \pi^2 f M^3_p R^2 w(\phi)^4 V_{f0}(\phi)^2}V^{\pm\, a}=0 \ ,
\end{aligned}
\end{equation}
which receive a contribution form the CS term.
These equations are actually diagonal in terms of the left and right handed fluctuations:
\begin{equation}
\begin{aligned}
\label{eq:LRinstabilityeq}
&(\delta A_L^{\pm\, a})''+\left[\frac{d}{dr} \log e^{A}f w(\phi)^2V_{f0}(\phi)R\right](\delta A_L^{\pm\, a})'\\
&+\left[\frac{\omega^2}{f^2}-\frac{q^2}{f R^2}\right]\delta A_L^{\pm\, a} \pm \frac{e^{-2 A} \hat{n} q }{2 \pi^2 f M^3_p R^2 w(\phi)^4 V_{f0}(\phi)^2}\delta A_L^{\pm\, a}=0
\end{aligned}
\end{equation}
and similarly for the right-handed field (but with opposite sign in the CS contribution). Here $\delta A_L^{\pm\,a}=\delta V^{\pm\,a}+\delta A^{\pm\,a}$.

The fluctuations in the Abelian sectors also couple to metric:
\begin{equation}
\begin{aligned}
&(Z_1^\pm)''+\frac{f A^{\prime}\left(\omega^2-f q^2\right)-\omega^2 f^{\prime}}{f\left(f q^2-\omega^2\right)}(Z_1^\pm)'+\left(\frac{2\omega^2 A'f'}{f(fq^2-\omega^2)}+\frac{\omega^2-fq^2}{f^2}+ \right.\\
&\left.\frac{8 \phi'}{9}-4A'^2\right)Z_1^\pm- e^{-A}\hat n q x(\delta \hat V^{\pm})'+ \frac{e^{-A} \hat n q x \omega^2 f^{\prime}}{f^2 q^2-f \omega^2}\delta \hat V^{\pm}=0
\end{aligned}
\end{equation}
\begin{equation}
\begin{aligned}
&(\delta \hat V^{\pm})''+\left[\frac{d}{dr} \log e^{A}f w(\phi)^2V_{f0}(\phi)R\right](\delta \hat V^{\pm})'\\
&+\left[\frac{\omega^2}{f^2}-\frac{q^2}{f R^2}-\frac{ 
x\hat n^2 \omega^2 e^{-4A}   }{\left(\omega^2 -f q^2\right) f V_{f0}(\phi) w(\phi)^2R}\right]\delta \hat V^{\pm}-\\
&\frac{\hat{n} q(Z_1^\pm)'}{w(\phi)\left(f q^2-\omega^2\right) e^{3A} R w(\phi) V_{f0}(\phi)}\pm\frac{e^{-2 A} \hat{n} q\delta \hat A^{\pm}}{2 \pi^2 f M^3_p R^2 w(\phi)^4 V_{f0}(\phi)^2}-\\
&\frac{2 \hat{n} q A^{\prime} Z_1^\pm}{w(\phi)\left(f q^2-\omega^2\right) e^{3A} R w(\phi) V_{f0}(\phi)}=0
\end{aligned}
\end{equation}
\begin{equation}
\begin{aligned}
&(\delta \hat A^{\pm})''+\left[\frac{d}{dr} \log e^{A}f w(\phi)^2V_{f0}(\phi)R\right](\delta \hat A^{\pm})'+\left[\frac{\omega^2}{f^2}-\frac{q^2}{f R^2}- \frac{4 x e^{2A}Z(\phi)}{f V_{f0}(\phi) w(\phi )^2 R} \right]\delta \hat A^{\pm}\\ 
&\pm \frac{e^{-2 A} \hat{n} q\delta \hat V^{\pm}}{2 \pi^2 f M^3_p R^2 w(\phi)^4 V_{f0}(\phi)^2}=0
\end{aligned}
\end{equation}
where
\be
R=\sqrt{1+\frac{\hat n^2}{e^{6 A} w(\phi)^2 V_{f0}(\phi)^2}} \ .
\ee

\subsection{Helicity zero}

The helicity zero sector is the most complicated sector. It further divides into several sets of coupled equations (see Fig.~\ref{finq_class}) which we discuss separately.

\paragraph{Equations for Abelian vector, helicity zero gravity, and the dilaton fluctuations.} These three fluctuation wave functions, $\hat E_V$, $Z_0$, and $Z_\phi$, satisfy a coupled system of three ordinary second order differential equations. We have derived this system but since it is extremely complicated (several pages) we do not include the expressions here. Note however that it passes the same consistency checks (listed above) as all the other equations.

\paragraph{Axion and Abelian axial vector equations.} These coupled equations can be written as
\begingroup
\allowdisplaybreaks
\begin{align}
\hat E_A'' &+ \Bigg[\frac{ 4 \hat n^2 q^2 x \omega ^2 e^{-6 A} Z(\phi )}{f w(\phi )^2 V_{f0}(\phi)^2 \left(q^2-\omega ^2/f\right) \left(q^2-R^2 \omega ^2/f\right)N_\mathrm{mix} }\,\frac{d }{d r}\log \frac{f}{R^2}+\nn\\
&+\frac{d }{d r}\log \frac{e^{A} w(\phi )^2 V_{f0}(\phi)R}{\omega ^2/f-q^2/R^2}\Bigg]\hat E_A'+ \Bigg[\frac{4 q^2 x \omega ^2 f' Z(\phi )}{f^2 \left(q^2-\omega ^2/f\right)^2N_\mathrm{mix} }\,\frac{d }{d r}\log \frac{f}{R^2} -\nn\\
&-\frac{\left(q^2-R^2 \omega ^2/f\right) \left(4 x e^{2 A} Z(\phi )+w(\phi )^2 V_{f0}(\phi) \left(q^2-\omega ^2/f\right)R\right)}{R^3 f w(\phi )^2 V_{f0}(\phi) \left(q^2-\omega ^2/f\right)}\Bigg]\hat E_A-\nn\\
&-\frac{2 i q \omega   Z(\phi )}{N_\mathrm{mix}}\,\frac{d }{d r}\log \frac{f}{R^2}\,Z_a'+\frac{ 2 i \hat n^2 q \omega  e^{-4 A} Z(\phi )}{ f w(\phi )^4 V_{f0}(\phi)^3R^3}\,Z_a = 0
\end{align}    
\begin{align}
Z_a'' &+\Bigg[\frac{d}{d r} \log \left(e^{3 A} f Z(\phi )\right)+\frac{4  x Z(\phi ) q^2}{ \left(q^2-\omega ^2/f\right)N_\mathrm{mix}}\,\frac{f'}{f}-\frac{4 x  Z(\phi )}{N_\mathrm{mix}}\,\frac{d}{dr}\log\frac{e^{5A}fZ(\phi)}{w(\phi)}-\nn\\
&-\frac{4 e^{6 A}  x \left(q^2+\left(2 R^2-3\right) \omega ^2/f\right) w(\phi )^2 V_{f0}(\phi)^2 Z(\phi )}{\hat n^2 N_\mathrm{mix} \left(q^2-\omega ^2/f\right)}\,\frac{R'}{R}\Bigg] Z_a' + \nn\\
&+\frac{\left(q^2-\omega ^2/f\right)^2 w(\phi )^2 V_{f0}(\phi) R^3+4 e^{2 A} x \left(q^2 R^2-\omega ^2/f\right) Z(\phi ) }{R^3 \left(\omega ^2/f-q^2\right) f w(\phi )^2 V_{f0}(\phi)}\,Z_a  +\nn\\
&+\Bigg[\frac{2 i e^{-2 A} x q \omega  \left(4 e^{2 A} x Z(\phi )-R \left(\left(R^2-2\right) q^2+R^2 \omega ^2/f\right) w(\phi )^2 V_{f0}(\phi)\right)}{f \left(q^2-\omega ^2/f\right)^2 N_\mathrm{mix}}\,\frac{f'}{f} +\nn\\
&+\frac{ 2 i e^{-2 A} x q \omega  \left(3 R \left(q^2-\omega ^2/f\right) V_{f0}(\phi) w(\phi )^2+8 e^{2 A} x Z(\phi )\right)}{f \left(q^2-\omega ^2/f\right)^2 N_\mathrm{mix}}\,\frac{R'}{R}+\nn\\
&+\frac{2 i e^{-8 A} \hat n^2 x q \omega R }{V_{f0}(\phi)\left(q^2 f-\omega ^2 \right)N_\mathrm{mix}}\,\frac{d}{dr}\log\frac{e^{5A}fZ(\phi)}{w(\phi)} \Bigg] \hat E_A' +\nn\\
&+ \Bigg[\frac{2 e^{-6 A} i \hat n^2 x q \omega  \left(R V_{f0}(\phi) \left(q^2+e^{2 A} R x V_{f0}(\phi)-\omega ^2/f\right) w(\phi )^2+4 e^{2 A} x Z(\phi )\right)}{R^3 \left(q^2-\omega ^2/f\right)^2 f^2 w(\phi )^4 V_{f0}(\phi)^3}-\nn\\
&-\frac{2 i x q \omega  \left(5 w(\phi ) A'-w'(\phi ) \phi '\right)}{\left(q^2-\omega ^2/f\right)^2 f w(\phi )}\,\frac{f'}{f}+\nn\\
&+\frac{4 i e^{-2 A} q^3 x \omega  \left(R \left(q^2-R^2 \omega ^2/f\right) V_{f0}(\phi) w(\phi )^2+2 e^{2 A} x Z(\phi )\right)}{ f \left(\omega ^2/f-q^2\right)^3 N_\mathrm{mix}}\left(\frac{f'}{f}\right)^2+\nn\\
&+\frac{ \left(8 i e^{6 A} q x^2 \omega  \left(q^2+\left(2 R^2-3\right) \omega ^2/f\right) w(\phi )^2 V_{f0}(\phi)^2 Z(\phi )\right)}{\hat n^2 f \left(\omega ^2/f-q^2\right)^3 N_\mathrm{mix}}\,\frac{f'}{f} \,\frac{R'}{R}+\nn\\
&+\frac{ 2 i e^{-2 A} q R x \omega  \left(q^2-R^2 \omega ^2/f\right) w(\phi )^2 V_{f0}(\phi)}{f \left(q^2-\omega ^2/f\right)^2 N_\mathrm{mix}}\,\frac{f'}{f}\, \frac{d}{dr}\log\frac{e^{5A}fZ(\phi)}{w(\phi)}\Bigg] \hat E_A = 0
\end{align}%
\endgroup
where\footnote{The point where $N_\mathrm{mix}$ vanishes gives rise to an apparent singularity of the fluctuation equations. It was shown in~\cite{Arean:2016hcs} that at zero temperature and density this singularity can be removed by a redefinition of the wave functions. We expect that the same is true here.}
\be \label{Nmixdef}
 N_\mathrm{mix} = 4 x  Z(\phi )+e^{-2 A} w(\phi )^2 V_{f0}(\phi ) \left(q^2-R^2 \omega ^2/f\right)R\ .
\ee

\paragraph{The tachyon equation.} The tachyon equation reads
\be
 \delta \tau''+ \frac{d}{dr}\left[\log e^{3 A} f \kappa (\phi) V_{f0}(\phi)R\right]\delta\tau' + \left[\frac{\omega ^2}{f^2}-\frac{q^2}{R^2f}+\frac{2 e^{2A}}{R^2f \kappa (\phi) V_{f0}(\phi)}\right] \delta \tau = 0 \ .
\ee

\paragraph{Non-Abelian gauge-field equations.} For the non-Abelian modes of the gauge fields we find the equation
\be
 \left(E_V^a\right)'' + \frac{d}{dr}\left[\log \frac{e^{A} V_{f0}(\phi ) w(\phi )^2 R}{ \omega ^2/f-q^2 /R^{2}}\right] \left(E_V^a\right)' + \left( \frac{\omega ^2}{f^2}-\frac{q^2}{fR^2} \right)E_V^a = 0
\ee
and the axial field $E_A^a$ satisfies the same equation. 

\subsection{Equations in the limit \texorpdfstring{$q \to 0$}{TEXT}}\label{app:q0flucts}

In the limit of zero momentum, the fluctuations can be classified according to their spin rather than helicity (or the $z$ component of the spin), as there is no longer preferred spatial direction, see Fig.~\ref{q0class}. All fluctuations are decoupled.\footnote{To be precise, the decoupling is not complete: in the helicity zero sector the fluctuation equation of $Z_\phi$ still includes dependence on $Z_2$ even when $q=0$. However, since these are the only mixing terms, they do not contribute to the QNMs. One way to see this to redefine $Z_2 = \sqrt{q}\, \tilde Z_2$ after which taking $q \to 0 $ leads to complete decoupling. }

\paragraph{Spin two.} All fluctuations in the spin two sector satisfy the equation of a massless probe scalar in the gravity background,
\be
F'' +\left(3 A'+\frac{f'}{f}\right)F' +\frac{\omega ^2 }{f^2}F = 0 \ ,
\ee
where $F = e^{-2A}Z_2^{\pm}$, $e^{-2A}Z_1^\pm$, $e^{-2A}Z_0$.

\paragraph{Spin one.} The non-Abelian gauge field fluctuations satisfy the equation
\be \label{NAbelq0eqs}
 F'' + \frac{d}{dr}\left[\log e^{A} f V_{f0}(\phi ) w(\phi )^2 R\right] F' +  \frac{\omega ^2}{f^2}F = 0 \ .
\ee
That is, here 
$F = E_V^a$, $\delta V^{a\pm}$, $E_A^a$, $\delta A^{a\pm}$. The equation for the Abelian vector and axial vector fields include additional mass terms:
\be 
\begin{aligned} \label{Abelq0eqs}
 F_V'' + \frac{d}{dr}\left[\log e^{A} f V_{f0}(\phi ) w(\phi )^2 R\right] F_V' +  \left(\frac{\omega ^2}{f^2}- \frac{\hat n^2 x e^{-4A}}{f V_{f0}(\phi) w(\phi )^2 R}\right)F_V &= 0 \ , \\
 F_A'' + \frac{d}{dr}\left[\log e^{A} f V_{f0}(\phi ) w(\phi )^2 R\right] F_A' +  \left(\frac{\omega ^2}{f^2}- \frac{4 x e^{2A}Z(\phi)}{f V_{f0}(\phi) w(\phi )^2 R}\right)F_A &= 0 \ ,
\end{aligned}
\ee
with $F_V = \hat E_V$, $\delta \hat V^\pm$ and $F_A = \hat E_A$, $\delta \hat A^\pm$. The vectorial mass term arises due to the coupling of the vectorial fluctuations to the vectorial gauge field in the background, and the axial mass term comes from the axion term in~\eqref{Sadef}.

\paragraph{Spin zero.} In the spin zero sector, we have the fluctuation equations for the dilaton, the tachyon, and the axion:
\begingroup
\allowdisplaybreaks
\begin{align} \label{Zphiq0eq}
 Z_\phi'' &+ \left(3 A' +\frac{f'}{f}\right)Z_\phi'+ \Bigg[\frac{\omega ^2}{f^2}+   & \\\nonumber
 &+ 
 \frac{e^{2A}}{f}\left(\frac{3}{8}\frac{\pa^2 V_\mathrm{eff}(\phi,A,\hat n)}{\pa \phi^2}+\frac{2\phi'}{3A'}\frac{\pa V_\mathrm{eff}(\phi,A,\hat n)}{\pa \phi}+\frac{8(\phi')^2}{27(A')^2}V_\mathrm{eff}(\phi,A,\hat n)\right)\Bigg]Z_\phi = 0 \\
   \delta \tau'' &+ \frac{d}{dr}\left[\log e^{3 A} f \kappa (\phi) V_{f0}(\phi)R\right]\delta\tau' + \left[\frac{\omega ^2}{f^2}+\frac{2 e^{2A}}{R^2f \kappa (\phi) V_{f0}(\phi)}\right] \delta \tau = 0 \\
  Z_a'' &+\Bigg[\frac{d}{d r} \log \left(e^{3 A} f Z(\phi )\right)-\frac{4 x  Z(\phi )}{N_\mathrm{mix}^{(0)}}\,\frac{d}{dr}\log\frac{e^{5A}fZ(\phi)}{w(\phi)}+\nn\\
&+\frac{4 e^{6 A}  x \left(2 R^2-3\right)  w(\phi )^2 V_{f0}(\phi)^2 Z(\phi )}{\hat n^2 N_\mathrm{mix}^{(0)}}\,\frac{R'}{R}\Bigg] Z_a' + \nn\\
&+\left[\frac{\omega^2}{f^2} -\frac{4 x e^{2 A} Z(\phi ) }{  f w(\phi )^2 V_{f0}(\phi)R^3}\right]\,Z_a = 0
\end{align}
\endgroup
where the effective potential $V_\mathrm{eff}$ was defined in~\eqref{Veffdef}, and at $q=0$, $N_\mathrm{mix}$ of~\eqref{Nmixdef} reduces to
\be
 N_\mathrm{mix}^{(0)} = 4 x  Z(\phi )-\omega ^2e^{-2 A} V_{f0}(\phi )w(\phi )^2  R^3 /f\ .
\ee

\subsection{Behavior of the dilaton equation near the AdS\texorpdfstring{$_2$}{TEXT} point} 

Finally, we discuss how the dilaton equation~\eqref{Zphiq0eq} behaves near the endpoint of the zero temperature flows that end in the AdS$_2$ geometry. The flow is given as an explicit asymptotic transseries expansion in~\cite{Hoyos:2021njg}. The terms relevant here can be written as
\be
 A = A_* + A_1 r^{2\alpha} + A_2 r + \cdots \ , \qquad \phi = \phi_* + \phi_1 r^\alpha + \cdots \ , \qquad f = f_0 r^2(1+\cdots) \ ,
\ee
where
\begin{align}
\label{eq:alphadef}
 \alpha &= \Delta_* - 1 = \frac{1}{2}\left[-1+\sqrt{1-\frac{9\, \partial_\phi^2 V_\mathrm{eff}(\phi_*,A_*,\hat n)}{\partial_A V_\mathrm{eff}(\phi_*,A_*,\hat n)}}\right]\ , & \\
 \label{eq:exprels}
 \frac{e^{2A_*}}{f_0} &= L_2^2 = \frac{6}{\partial_A V_\mathrm{eff}(\phi_*,A_*,\hat n)} \ , \qquad \frac{\phi_1^2}{A_1} = \frac{9(1-2\alpha)}{2\alpha} \ . &
\end{align}
Here we already assumed that $0<\alpha<1$. This parameter is always positive, and if $\alpha>1$, it is immediate that the equation reduces to that given in~\eqref{dilads2eq}. For the coefficient of the first order derivative we find for all values of $\alpha$ that
\be
3 A' +\frac{f'}{f} \approx \frac{2}{r} \ .
\ee
Let us then assume that $1/2<\alpha<1$. In this case
\begin{align}
\frac{\phi'}{A'} &\approx \frac{\alpha \phi_1 r^{\alpha-1} }{A_2} \ , \qquad \partial_\phi V_\mathrm{eff}(\phi,A,\hat n) \approx \phi_1 r^\alpha\, \partial_\phi^2 V_\mathrm{eff}(\phi_*,A_*,\hat n) \ , & \\
 V_\mathrm{eff}(\phi,A,\hat n) &\approx A_2 r\, \partial_A V_\mathrm{eff}(\phi_*,A_*,\hat n)\ , &
\end{align}
where we used the fact that both the effective potential and its first $\phi$-derivative vanish at the fixed point.
Inserting these in the mass term on the second row of~\eqref{Zphiq0eq}, we see that the first factor dominates, so that~\eqref{dilads2eq} is again reproduced.

However, if $0<\alpha<1/2$ we find instead
\begin{align}
\frac{\phi'}{A'} &\approx \frac{\phi_1  }{A_1 r^{\alpha}} \ , \qquad \partial_\phi V_\mathrm{eff}(\phi,A,\hat n) \approx \phi_1 r^\alpha\, \partial_\phi^2 V_\mathrm{eff}(\phi_*,A_*,\hat n) \ , & \\
V_\mathrm{eff}(\phi,A,\hat n) &\approx \frac{1}{2}\phi_1^2r^{2\alpha}\partial_\phi^2 V_\mathrm{eff}(\phi_*,A_*,\hat n) +A_1 r^{2\alpha}\, \partial_A V_\mathrm{eff}(\phi_*,A_*,\hat n) \ . &
\end{align}
Now all three factors in the mass term contribute. After some algebra and using the relations in~\eqref{eq:alphadef} and in~\eqref{eq:exprels}, we find that
\begin{align}
    \frac{e^{2A}}{f}\!\left(\frac{3}{8}\frac{\pa^2 V_\mathrm{eff}(\phi,A,\hat n)}{\pa \phi^2}+\frac{2\phi'}{3A'}\frac{\pa V_\mathrm{eff}(\phi,A,\hat n)}{\pa \phi}+\frac{8(\phi')^2}{27(A')^2}V_\mathrm{eff}(\phi,A,\hat n)\right) \approx \frac{-\alpha ^2+3 \alpha -2}{r^2} \ .
\end{align}
The solutions of the resulting equation (at zero frequency)
\be
 Z_\phi''+\frac{2}{r} Z_\phi' + \frac{-\alpha ^2+3 \alpha -2}{r^2} Z_\phi = 0
\ee
are given by
\be
 Z_\phi  
 =C_1 r^{\Delta_*^\mathrm{fl}-1} +C_2 r^{-\Delta_*^\mathrm{fl}} 
\ee
with 
\be
 \Delta_*^\mathrm{fl} = 2-\alpha = 3-\Delta_* = \frac{1}{2}\left[5-\sqrt{1-\frac{9\, \partial_\phi^2 V_\mathrm{eff}(\phi_*,A_*,\hat n)}{\partial_A V_\mathrm{eff}(\phi_*,A_*,\hat n)}}\right] \ .
\ee
Recall that this result holds for $0<\alpha<1/2$. Combining this with the result for the case of $\alpha>1/2$ (so that $\Delta_*>3/2$), we recover~\eqref{dildeltaflow}.

\section{Numerical method} \label{app:numerics}

 In order to determine QNM spectrum, one needs to solve numerically a coupled system ordinary differential equations (ODEs) that is described in the previous Appendix \ref{flucEq}.
 We use pseudospectral methods,  a well-established approach frequently employed  in gravity \cite{Dias:2009iu,Dias:2010eu,Monteiro:2009ke} and in holography \cite{Gursoy:2016ggq,Jansen:2017oag,Bosch:2017ccw,Gursoy:2016ggq,Buchel:2017map,Bu:2019mow,Gursoy:2016tgf} for QNM computations.  In this appendix, we provide a concise overview of this method (see, e.g., \cite{boyd01} for more details) and how it can be  applied to the specific equations derived in this article. 

We start by discussing the choice of coordinates. Eddington-Finkelstein (EF) coordinates are well-suited for implementing the horizon boundary conditions when using pseudospectral methods. 
 In EF coordinates the metric in (\ref{metric}) is written as
 \begin{equation} \label{metricapp}
 ds^2=e^{2A(r)}\left[-f(r)(dt')^2+2
 dt'dr+d\mathbf{x}^2\right], 
 \end{equation}
 where $t'$ is EF time coordinate defined through  
 the exact differential $dt' = dt + f(r)^{-1} dr$. This definition directly integrates to $t' = t  + G(r)$, where $G'(r) = f(r)^{-1}$. Recalling that the time dependence of the fluctuation is that of a plane wave, $\propto e^{-i \omega t}$, this implies that the fluctuation wave functions $\psi$ transform as $\psi\mapsto e^{i\omega G(r)}\psi$.
 Near the horizon, all fluctuation wave functions can be expressed as a sum of two solutions: 
an ingoing solution with a constant leading term,  
and an outgoing solution that oscillates rapidly. 
This is convenient when using the pseudospectral method, because the method automatically picks the regular solution, i.e., the ingoing solution, which is the horizon boundary condition dictated by causality.

It is also convenient to carry out a change of coordinates in the holographic direction, first from the conformal coordinate $r$   
to $A$ given in~\eqref{metricapp}. In these coordinates the boundary is located at $A=+\infty$. Therefore we impose a UV cutoff and only solve the fluctuations in a finite range of $A$. The fluctuation wave functions typically decrease exponentially with $A$ so a moderate cutoff at $A=\mathcal{O}(10)$ is enough. As a second coordinate transformation, we apply a linear mapping from $A$ to a new coordinate, defined in the interval $[-1,1]$, which is the standard interval for the pseudospectral method.

Finally, we need to choose the normalizable UV boundary condition for the fluctuations. In order to implement this, we need to make sure that the independent nonnormalizable solution diverges at the boundary. If it goes to a constant or vanishes, we apply a redefinition of the wave function (e.g., by an exponential factor in $A$) such that the normalizable solution is the only solution remaining finite at the boundary. After this, the pseudospectral method, which can only produce regular solutions to differential equations, automatically excludes the nonnormalizable solution.

 The essence of pseudospectral method is the discretization of differential equations to turn them into 
 matrix equations. This procedure involves replacing the continuous independent variable with a discrete counterpart. The domain of this discrete variable commonly referred to as the collocation grid. 
 Functions can then be approximated by 
 \begin{align}\label{chebex}
   f(x)\approx\sum_{i=0}^N f(x_i)C_i(x) \ ,
 \end{align}
 where $x_i$ denotes the collocation grid points $\{x_i|i=0,\dots, N\}$ and $C_i(x)$ are 
 the cardinal 
 functions satisfying $C_i(x_j)=\delta_{ij}$ so that the approximation~\eqref{chebex} is exact on the grid points.
 This expansion enables one to estimate the derivatives of the function $f$ 
 by using derivative matrices constructed from the derivatives of the cardinal functions on the grid points. That is, 
 \be \label{derappr}
 f^{(n)}(x_i) \approx \sum_j D_{ij}^{(n)} f(x_j)
 \ee 
 where $D_{ij}^{(n)}=C_j^{(n)}(x_i)$. For $n=0$, we take $D_{ij}^{(0)}=\delta_{ij}$. 

 The standard choice for the grid, which we also use, is the Chebyshev–Gauss–Lobatto (CGL) grid. The definition on the interval $[-1,1]$ is
 \begin{equation}
     x_i=\cos \left(\frac{\pi i}{N}\right), \qquad i=0,1,\dots, N\ .
 \end{equation}
The cardinal functions are constructed from Chebyshev polynomials $T_n$ as
\begin{equation}
    C_i(x)=\frac{2}{Np_i}\sum_{n=0}^N \frac{1}{p_n}T_n(x_i)T_n(x) \ ,
\end{equation}
where $p_0=p_N=2$ and $p_j=1$ for $0<j<N$.
One can show that for this choice of the grid the error in the approximation (\ref{chebex}) decreases exponentially with increasing grid size $N$.

Let us consider for simplicity a fluctuation equation for one decoupled wave function $\psi$ at zero momentum. It can be written as 

\begin{equation}\label{eigeneq}
M(\omega)\psi=
\sum_{I=0}^p \omega^I M_I\psi=0,
\end{equation}
where  $p$ is the highest power of the frequency $\omega$ appearing in the equation, and $M(\omega)$ and $M_I$ are differential operators involving derivatives of the holographic coordinate, and the functions forming the background solution. The idea is then to evaluate the equation on the collocation points, and apply the approximation~\eqref{derappr} to the differential operators. The resulting $N+1$ equations only depend on $f$ through its values on the grid points. That is, the operators $M_I$ are replaced by the matrices 
\begin{equation}
        (\widetilde M_I)_{ij}=\sum_{n=0}^2 c_{nI}(x_j)D^{(n)}_{ij}\ ,
\end{equation}
where the coefficients $c_{nI}(x_j)$ are obtained by inserting the chosen background evaluated at the collocation points. Similarly, the wave function is replaced by the vector $\tilde \psi$, obtained by discretizing the function as $\tilde \psi_j = \psi(x_j)$. Defining $\widetilde M (\omega) = \sum_{I=0}^p \omega^I \widetilde M_I$, we can write the final discretized counterpart of~\eqref{eigeneq}:
\be \label{finalmatrixeq}
 \widetilde M(\omega) \widetilde\psi = 0 \ .
\ee
The QNMs are then approximated by nontrivial solutions to this equation, which occur when the determinant of the matrix vanishes. In order to locate the QNMs, one therefore just needs to find numerically the roots of the equation $\det(\widetilde M(\omega))=0$.

We test the convergence of the modes by varying the grid size $N$. 
We find that a larger grid is necessary for convergence at low temperature values,  near the AdS$_2$ critical point. At high temperatures $N=50$ is sufficient to compute the locations of the leading modes, but for the lowest temperatures considered in this articles, $N=500$ or even more is required for convergence. 

Finally, 
 as a nontrivial consistency check, the QNMs behave as expected in the limit of zero momentum. In particular, the AdS$_2$ modes at low temperatures are in agreement with analytical computations, and the diffusive, shear, and sound hydro modes show the expected 
dispersion relation in the limit $q\rightarrow 0$.

\bibliographystyle{JHEP}
\bibliography{refs} 

\end{document}